\documentclass[hidelinks,12pt]{article}
\usepackage[utf8]{inputenc}
\usepackage[english]{babel}
\usepackage[a4paper,lmargin={1.99cm},rmargin={1.99cm},tmargin={2.5cm},bmargin={2.5cm}]{geometry}
\usepackage{color}
\usepackage{mathptmx}
\usepackage{subcaption}
\usepackage{amssymb}
\usepackage{amsmath,amsfonts,amsthm,bm}
\usepackage{graphicx, multicol}
\usepackage{epstopdf}
\usepackage[onehalfspacing]{setspace}
\usepackage{nth}
\usepackage[dvipsnames]{xcolor}
\usepackage{qtree}
\usepackage{verbatim}
\usepackage{csquotes}
\usepackage{threeparttable}

\usepackage{pdflscape}
\usepackage{afterpage}

\usepackage{tabularx}
\usepackage{array}
\newcolumntype{b}{X}
\newcolumntype{s}{>{\hsize=.5\hsize}X}
\newcolumntype{Y}{>{\hsize=11.5\hsize}X}
\newcolumntype{G}{>{\centering\arraybackslash\hsize=4.25\hsize}X}

\usepackage{mathpazo} 
\usepackage[scaled]{helvet} 
\usepackage{courier} 
\normalfont
\usepackage[T1]{fontenc}
\usepackage{listings}
\usepackage{colortbl}
\definecolor{siena}{rgb}{0.91,0.45,0.32} 	

\usepackage[bookmarks,colorlinks]{hyperref}
\hypersetup{
  colorlinks,
  citecolor=PineGreen,
  linkcolor=red,
  urlcolor=blue}
\usepackage{cleveref}

\usepackage{float}
\usepackage{booktabs}
\usepackage{titling}
\usepackage{natbib}
\usepackage{bibunits}

\thanksmarkseries{arabic}
\begin{document}
\vspace{-3cm}

\title{\Huge Arctic Amplification of Anthropogenic Forcing: \\ A Vector Autoregressive Analysis}
\author{\hspace{-0.75cm} Philippe Goulet Coulombe\thanks{%
Department of Economics,  \href{mailto:gouletc@sas.upenn.edu}{{gouletc@sas.upenn.edu}}. For helpful discussions and comments on earlier drafts, we would like to thank Edvard Bakhitov, Elizaveta Brover, Francesco Corsello, Frank Diebold, Marie McGraw, Tony Liu, Glenn Rudebusch, Dalibor Stevanovic, David Wigglesworth, Boyuan Zhang
and members of the Penn Climate Econometrics Group. Moreover, we are thankful to 3 anonymous referees whose comments and suggestions greatly ameliorated this paper.} \qquad \qquad \hspace{0.15cm} Maximilian G\"obel}
\date{\vspace{-0.4cm} \hspace{0.95cm} University of Pennsylvania \qquad \hspace{-0.1cm} ISEG - Universidade de Lisboa  \\[2ex]%
\small
First Draft: December 11, 2019 \\
This Draft: January 17, 2021 \\ 
\vspace{0.4cm}
\large
 } 

\newgeometry{left=1.3cm, right = 1.3cm, bottom = 2.0cm, top = 2.0cm}
\maketitle\thispagestyle{empty}
\begin{abstract}

\noindent On September 15$^\text{th} $ 2020, Arctic sea ice extent (SIE) ranked second-to-lowest in history and keeps trending downward. The understanding of how feedback loops amplify the effects of external $CO_2$ forcing is still limited. We propose the VARCTIC, which is a Vector Autoregression (VAR) designed to capture and extrapolate Arctic feedback loops. VARs are dynamic simultaneous systems of equations, routinely estimated to predict and understand the interactions of multiple macroeconomic time series. The VARCTIC is a parsimonious compromise between full-blown climate models and purely statistical approaches that usually offer little explanation of the underlying mechanism. 
Our completely unconditional forecast has SIE hitting 0 in September by the 2060's. 
Impulse response functions reveal  that anthropogenic $CO_2$ emission shocks have an unusually durable effect on SIE -- a property shared by no other shock. We find Albedo- and Thickness-based feedbacks to be the main amplification channels through which $CO_2$ anomalies impact SIE in the short/medium run. 
Further, conditional forecast analyses reveal that the future path of SIE crucially depends on the evolution of $CO_2$ emissions, with outcomes ranging from recovering SIE to it reaching 0 in the 2050's. Finally, Albedo and Thickness feedbacks are shown to play an important role in accelerating the speed at which predicted SIE is heading towards 0.

\end{abstract}

\restoregeometry




\clearpage


\clearpage

\newpage
\pagenumbering{arabic}
\section{Introduction} \label{Intro}

With 3.74 million square kilometers on September 15$^\text{th}$ 2020, Arctic sea ice extent ranked second-to-lowest in history, after the record minimum in 2012. A persistent retreat of SIE may further accelerate global warming and threaten the composition of the Arctic's ecosystem (\cite{ScSim10}). The Coupled Model Intercomparison Project (CMIP) assembles estimates of long-run projections of Arctic sea ice from many climate models. These models try to reproduce the geophysical dynamics and interrelations among various variables, influencing the evolution of global climate. 

With CMIP being in its 6$^{\text{th}}$ phase (CMIP6), climate models now provide more realistic forecasts of the Arctic's sea ice cover compared to its predecessor CMIP5 (see \cite{StroeveEtAl12}, \cite{Notzetal2020}). The majority of contributors to CMIP6 see the Arctic's September mean sea ice to retreat below the $1 \times 10^6$ $km^2$ mark before the year 2050. Despite following the hitherto accepted physical laws of our climate, its chaotic nature, i.e. the still obscure interplay of various climate variables, imposes a major burden on climate models. 
Repeated initialization with differing starting conditions is intended to reduce uncertainty and biases surrounding initial parameters. The byproduct is a wide range of projections of key climate variables \citep{Notzetal2020}. In addition to such tuning, these simulations require large amounts of input data and a coupling of various sub-models \citep{TaylorEtAl09}. 

The above raises the question whether an approach that is statistical and yet \textit{multivariate} 
can paint a more conciliating picture. This means estimating a statistical system that depicts the interaction of key variables describing the state of the Arctic. In such a setup, the downward SIE path will be an implication of a complete dynamic system based on the observed climate record.
We provide a formal statistical assessment of different hypotheses about the historical path of SIE and outline the implications for the future. The effects on Arctic sea ice arising from various physical processes -- and the uncertainty surrounding their estimation -- can both be quantified without resorting to use a climate model.
\vskip 0.2cm

{\sc \noindent \textbf{Feedback Loops}.}  \emph{Feedbacks} are dynamics initially triggered by an \emph{external shock} to the system. Such a disturbance can be of radiative nature or not.\footnote{In contrast, \emph{internal variability}, another source of climactic variation, describes fluctuations emerging from within the climate system  \citep{KayEtAl2015}.}
Our analysis aims at better understanding how feedback loops -- and their interaction with anthropogenic carbon dioxide ($CO_2$) forcing -- shape the response of key Arctic variables, and most notably, sea ice.\footnote{A detailed description of various feedbacks, which the VARCTIC is capable of modeling, can be found in \citep{GoosseEtAl2018}.}
$CO_2$ forcing is already widely suspected to be the main driver behind long-run SIE evolution (see \cite{Meier2014}, \cite{notz2017arctic}). 
Feedback loops are well documented in the literature (see 
\cite{Park13}, \cite{Winton13}, \cite{StueEtAl18}, \cite{McGraw19}) and their understanding is crucial for enhancing the predictability of the Arctic's sea ice cover (\cite{Wang16}, \cite{Notzetal2020}). 
Only an approach that considers the interaction of many variables in a flexible way -- and thus numerous potential sources for feedback loops -- has a chance to depict a reliable statistical portrait of the Arctic. CMIP6 models consider many variables, but high variation in sea ice projections (see \cite{Notzetal2020}) suggests (among other things) widespread uncertainty around how strongly feedback loops may amplify external forcing. To shed more compelling statistical light on the matter, we borrow a methodology from economics. 

\vskip 0.2cm

{\sc \noindent \textbf{The Varctic}.} Our analysis focuses on the evolution of the long-term trajectory of SIE and the interdependent processes behind it. The modeling approach, which we propose, achieves a desirable balance between purely statistical and theoretical/structural approaches. In many fields, statistical approaches often have a better forecasting record than theory-based models.\footnote{ When it comes to September Arctic sea ice, statistical approaches supplanted dynamical models for at least the last three years, as per the Sea Ice Prediction Network's Sea Ice Outlook post-season reports \citep{SIOAug2019}. Statistical models showed much less disparity than theory-based alternatives and, most importantly, consistently provided a median forecast closer to the realized value.}
An obvious drawback is that the successful statistical model may provide little to no explanation of the underlying physical processes.

A Vector Autoregression (VAR) lives in a useful middle ground. It is a statistical model that yet generates forecasts by iterating a complete system of difference equations in multiple endogenous variables. These interactions can be analyzed and provide an explanation for the resulting forecasts. Considering all this, we propose the VAR for the Arctic (VARCTIC), a statistical approach that \textbf{(i)} can generate long-run forecasts, \textbf{(ii)} can explain them as the result of feedback loops and external forcing \textbf{(iii)} allows us to analyze how the Arctic responds to exogenous impulses/anomalies. 


\vskip 0.2cm

{\sc \noindent \textbf{Roadmap}.} We first discuss the data and its transformation in section \ref{sec:data}. We proceed with discussing the VAR model, its identification and Bayesian estimation in section \ref{sec:estimation}. Section \ref{sec:results} contains the empirical results which comprise (i) a long-run forecast of SIE, (ii) impulse response functions of the VAR, (iii) an exploration of the transmission mechanism (feedback loops), and (iv) a conditional forecasting analysis. We conclude and propose directions for future research in section \ref{conclusion}.

\section{Data} \label{sec:data}

Our data set comprises eighteen time series, proxying the Arctic's climate system, and accounting for potential feedback loops among the different constituents. The sample covers monthly observations from 1980 through 2018. We rely on standard data providers (see \cite{stroeve2018}), which are listed in Table \ref{tab:ListofAbb} in the appendix.
We combine eight variables, which importance has been highlighted by the existing literature (\cite{Meier2014}), into VARCTIC 8, our benchmark specification. Fortunately, variables can easily be added/removed from a VAR. Bayesian shrinkage ensures that a larger model will not overfit -- the latter aspect is further explained in section \ref{sec:bayes}. Therefore, the appendix contains VARCTIC 18 which includes an additional 10 series from the \emph{reanalysis product MERRA2} (\cite{MERRA2}) as a robustness check.
To summarize compactly, the two specifications considered in this paper are:  
\begin{itemize}
	\item[I]{\textbf{VARCTIC 8}}: $CO_2$, Total Cloud Cover (TCC), Precipitation Rate (PR), Air Temperature (AT), Sea Surface Temperature (SST), Sea Ice Extent (SIE), Sea Ice Thickness (SIT),  Albedo;
	\item[II.]{\textbf{VARCTIC 18}}: SWGNT, SWTNT, $CO_2$, LWGNT, TCC, TAUTOT, PR, TS, AT, SST, LWGAB, LWTUP, LWGEM, SIE, Age, SIT, EMIS, Albedo.
\end{itemize}
A comprehensive overview of all variables (including those of VARCTIC 18), their acronyms, and links to data providers can be found in the appendix in Table \ref{tab:ListofAbb}.\footnote{The primary goal was to assemble \textit{empirical} data on key climate variables. To capture the most prominent feedbacks on SIE (see \cite{Meier2014}), we augmented the observed series for $CO_2$, SIE, PR, and the assimilated PIOMAS product SIT, with data from model output. Our choice of data series is conditioned on whether they are (i) operated by well-established climate science institutions (ii) cited in the literature.
} We want the VARCTIC to be a credible approximation of a completely endogenous system, where local processes jointly determine each other, without significant external dependencies outside of forcing.\footnote{This is precisely what allows us to iterate the system forward (see section \ref{sec:estimation}\ref{sec:fcast}) in order to obtain statistical forecasts based on a dynamic system.} Thus, we restrict the spatial coverage to a regional rather than a global scale. In line with the literature \citep{notz2016observed}, all variables (except $CO_2$, {which is measured globally, and SST, which is measured over the Northern Hemisphere \citep{HorvathEtAl2020}}) are monthly means over all grid-cells between 30$^{\circ}$N and 90$^{\circ}$N latitude. {This region matches the spatial coverage of the Sea Ice Index and is in the neighborhood of the lower bound of 40$^{\circ}$N latitude applied in \cite{HorvathEtAl2020}.\footnote{Previous studies have emphasized the interdependencies between weather effects in the midlatitudes and the Arctic (\citealt{mcgrawbarnes2018,ScreenEtAl2015}).} It is a legitimate concern that averaging over too large of a region could wrongfully blend together mid-latitude events with others more specific to the Arctic circle. Fortunately, all key findings remain unchanged when restricting the gridded variables of TCC, PR, AT, and Albedo to the 60$^{\circ}$N-90$^{\circ}$N domain. An interesting avenue for future research is to consider a (larger) VAR where means over various latitude ranges are included -- so to study their interactions and relationship with SIE.}
Further, we follow \cite{OelkeEtAl2003} and use AT measured at a sigma-level of 0.995, i.e. at 0.995 $times$ each grid-cell's surface-level pressure. For its part, the important choice of VARCTIC 8's variables themselves (and additions in VARCTIC 18) will be motivated extensively in section \ref{sec:ID}.

The raw data is highly seasonal --- but the feedback loops we wish to estimate and extrapolate, reside in the (stochastic) trend components and short-run anomalies. Hence, we proceed to transform the data so that the resulting VARCTIC is fitted on deviations from seasonal means. 
For our benchmark analysis, we use a simple and transparent transformation: we de-seasonalize our data by regressing a particular variable $y^{raw}$ on a set of monthly dummies. That is, for each variable we run the regression
\begin{align}
y_{t}^{raw} = \sum_{m=1}^{12}\alpha_m D_m + \text{residual}_t
\end{align} 
with $y_t$ being defined as $y_t \equiv y_{t}^{raw} - \sum_{m=1}^{12}\hat{\alpha}_m D_m$. $D_m$ is an indicator that is 1 if date $t$ is in month $m$ and 0 otherwise. The estimates of $\alpha_m$, $\hat{\alpha}_m$, are obtained by ordinary least squares. This is exactly equivalent to de-meaning each data series month by month and is a more flexible approach to modeling seasonality than using Fourier series.\footnote{Of course, if we were using higher-frequency data -- like daily observations, then the Fourier approach would be much more parsimonious and potentially preferable \citep{hyndman2010forecasting}. The dummies approach to taking out seasonality only requires 12 coefficients with monthly, but 365 with hypothetical daily data.} Finally, we keep our filtered data $y$ in levels. We do not want to employ first differences or growth rate transformations to make the data stationary. Such an action would suppress long-run relationships which are an important object of interest. Figure \ref{fig:20Vars_ds} in the Appendix shows the data after being filtered with monthly dummies.\footnote{Note that $CO_2$ is available without seasonality from the data provider (NOAA-ESRL) and thus was not passed through the dummies filter.}

\begin{figure}[h!]
\centering
\includegraphics[width=\textwidth]{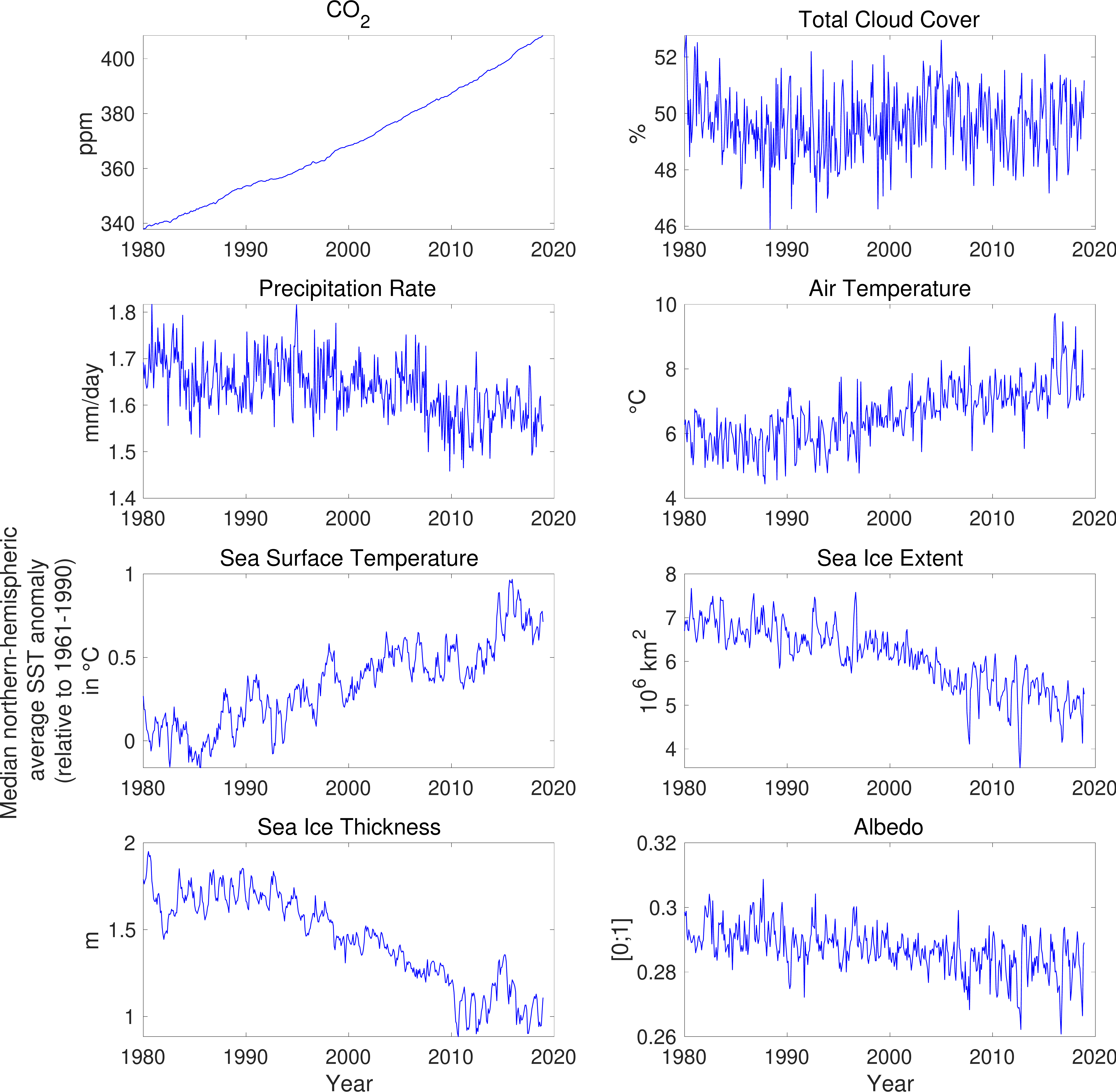}\\
\vspace*{0cm}\caption{Deseasonalized Series: 8 Variables}\label{fig:20Vars_ds}
\end{figure}

\textcolor{black}{Pre-processing the data can influence results. Moreover, \cite{DieRu19} and \cite{Meier2014} document seasonal variability in SIE trends. As a natural robustness check, we also consider a very different approach to eliminate seasonality. In appendix \ref{sec:ssm_theory}, we reproduce our results with a data set of stochastically de-seasonalized variables obtained from the approach of structural time series (\cite{Harvey90} and \cite{harvey2014}). In short, this extension allows for seasonality to evolve (slowly) over time, which could be a feature of some Arctic time series.}

\section{The VARCTIC}\label{sec:estimation}
In this section, we review the VAR: the model; its identification; its Bayesian estimation. Furthermore, we discuss the construction of the long-run forecasts and impulse response functions as tools to understand the VARCTICs' results.

\subsection{Vector Autoregressions and Climate}
Vector Autoregressions are dynamic simultaneous systems of equations. They can characterize a linear dynamic system in discrete time. The methodology was introduced to macroeconomics by \cite{sims1980} and is now so widely used that it almost became a field of its own (see \cite{kilian2017svar}). It is a multivariate model in the sense that $\boldsymbol{y}_t$ in
\begin{align}\label{struct_VAR}
A\boldsymbol{y}_t = \Psi_0 + \sum_{p=1}^{P}{\Psi_p}\boldsymbol{y}_{t-p} + \boldsymbol{\varepsilon}_t  ,
\end{align}
is an $M$ by 1 vector. This means that the dynamic system incorporates $M$ variables. $\Psi_p$'s parameterizes how each of these variables is predicted by its own lags and lags of the $M-1$ remaining variables. The matrix $A$ characterizes how the $M$ different variables interact contemporaneously --- e.g., how AT affects SIE within the \textit{same} month (a time unit $t$ in our setup). Finally, the disturbances are mutually uncorrelated with mean zero:	
\[
\boldsymbol{\varepsilon}_t = \left[ \varepsilon_{1,t},\enskip ... \enskip , \varepsilon_{M,t} \right]  
~ \sim ~ \, N \left (0 , ~ I_M \right ).
\]
Equation (\ref{struct_VAR}) is the so-called structural form of the VAR, which cannot be estimated because $A$ is not identified by the data. For clarity, the elements of $A$ are not plain regression coefficients, but structural model parameters. Attempting to estimate those directly via least squares would be plagued by simultaneity bias \citep{kilian2017svar}. Rather, \textit{structural} VAR estimation proceeds in two steps. First, an estimable "reduced-form" VAR is fitted to the data. That is, we run
\begin{align}\label{rf_VAR}
\boldsymbol{y}_t = \boldsymbol{c} + \sum_{p=1}^{P}{\Phi_p}\boldsymbol{y}_{t-p} + \boldsymbol{u}_t  ,
\end{align}
where $\boldsymbol{c} =  A^{-1} \Psi_0$ and $\Phi_p = A^{-1} \Psi_p$ are both regression coefficients. $\boldsymbol{u}_t$ are now regression residuals
\[
\boldsymbol{u}_t = \left[ u_{1,t},\enskip ... \enskip , u_{M,t} \right]  
~ \sim ~ \, N \left (0 , ~ \Sigma_u \right )
\]
which are allowed to be cross-correlated. By construction, $\Sigma_u=A^{-1'}A^{-1}$. This last relationship is key to the second, so-called "identification", step. In words, this means the covariance matrix of regression residuals from the first step ($\Sigma_u$) can be used as raw material to retrieve the "structural" $A$ --- the latter which, as we stressed earlier, cannot be estimated directly. The process for obtaining $A$ by decomposing $\Sigma_u$ is addressed on its own in section \ref{sec:ID}. 

The methodology has many advantages over simple autoregressive distributed lags (ARDL) regression that have gained some popularity in the econometric and climate literature. For instance, in \cite{mcgrawbarnes2018}, the argument for inclusion of lags of the dependent variable can be interpreted as one for completeness of the modeled dynamic system, as guaranteed by an adequately specified VAR. 

\subsection{Obtaining Long-Run Forecasts from a VAR}\label{sec:fcast}
The symmetry of the VAR allows for it to generate forecasts by simply iterating the model.\footnote{Further, forecasting does not rely on the matrix $A$.} Assuming the chosen variables to characterize the system completely, we can forecast its future state by iterating a particular mapping. To do so, we use a representation that exploits the fact that any VAR($P$) (that is, with $P$ lags) can be rewritten as a VAR(1), using the so-called companion matrix.\footnote{In short, any VAR($P$) in $M$ variables can be rewritten as a VAR(1) in $M \times P$ variables, such that the theoretical analysis can be carried out with the less burdensome VAR(1) \citep{kilian2017svar}.  $\boldsymbol{Y}_t$ are stacked $\boldsymbol{y}_{t-p}$'s for $p=1,...,P$.} Thus, obtaining forecasts amounts to iterate
\begin{align}\label{eq:it_fcst}
 \hat{\boldsymbol{Y}}_{t+1}=F(\hat{\boldsymbol{Y}}_t) \equiv \boldsymbol{\kappa} + {\Phi}\hat{\boldsymbol{Y}}_{t}, \quad \text{to obtain} \quad \hat{\boldsymbol{Y}}_{t+h} = F^h(\boldsymbol{Y}_{t}).
 \end{align}
\noindent where $F$ is the one-month ahead forecasting function, while $\boldsymbol{\kappa}$ and $\Phi$ are the companion-form analogs of $c$ and $\Phi_p$'s in \eqref{rf_VAR}. This equation provides forecasts of all variables, $h$ periods from time $t$. An obvious $t$ to consider is $T$, the end of the sample. The fact that we can obtain predictions by simply iterating the system, is of interest to generate scenarios for the Arctic. First, the prediction will rely on an explainable mechanism -- potentially mixing external forcing and internal feedback loops -- rather than a purely statistical relationship. Second, our forecast does not rely in any way on external data or forecasts made exogenously by some other entity, which would rely on assumptions implicitly incompatible with ours. Nevertheless, in some cases, it may be desirable to mix some external forecasts/scenarios of certain variables (like $CO_2$) that may be less successfully characterized by the VAR. We consider just that in section \ref{policy}. 
\subsection{Identification}\label{sec:ID}

While conditional and unconditional forecasting are important byproducts of the VARCTIC, another objective of our analysis is to understand -- from a statistical standpoint -- the underlying process driving interactions between key Arctic variables. For instance, by forecasting SIE \textit{conditional} on various emission scenarios, we will later show that anthropogenic $CO_2$ forcing is the main driving force behind the long-run forecast --- cutting emissions dramatically would prevent SIE from going to 0.\footnote{In contrast, an \textit{unconditional} forecast lets the VARCTIC generate internally future paths for all variables (including $CO_2$) without relying on externally developed scenarios.} This important result rests solely on the reduced-form VAR. However, to uncover and interpret the mechanism that amplifies $CO_2$'s effect on SIE, we need an identification scheme for instantaneous relationships. In time series analysis, the identification problem originates from simultaneity in the data. Multivariate time series data can tell us whether 
$X_{t-1} \rightarrow Y_{t} \quad \text{or} \quad Y_{t-1} \rightarrow X_{t}$
is more plausible. This is predictive causality in the sense of \cite{granger1969}. However, the data by itself cannot distinguish 
$X_{t} \rightarrow Y_{t} \quad \text{from} \quad X_{t} \leftarrow Y_{t}.$
In words, a single correlation between $X_t$ and $Y_t$ can be generated by two different causal structures. Within the VAR, the problem boils down to the need for identifying $A$ in equation (\ref{struct_VAR}). Yet, the data only procures us with the variance-covariance matrix of the residuals $\hat{\Sigma}$. The identification problem emerges from the fact that $A$ is not the only matrix satisfying $\hat{\Sigma}_u=A^{-1'}A^{-1}$.  Fortunately, there exist many ways to pin down a single $A$ matrix without having to delve into too much theory, which is partially responsible for the popularity of VARs among applied economists. The strategy we opt for is the traditional Choleski decomposition of $\hat{\Sigma}_u$. Mechanically, it provides a lower-triangular matrix $C$, satisfying $\hat{\Sigma}_u=C'C$ (where $C\equiv A^{-1}$ for convenience). Its purpose is to transform regressions residuals $u_t$ (equation \eqref{rf_VAR}) into uncorrelated structural shocks $\boldsymbol{\varepsilon}_t$ (equation \eqref{struct_VAR}). This is done by reversing the relationship $\boldsymbol{u}_t=C\boldsymbol{\varepsilon}_t$.
Uncorrelatedness is essential (as further discussed in section \ref{sec:irf}) to study how the VARCTIC responds to a given impulse, \textit{keeping everything else constant}.
Such a causal claim would be impossible when considering an impulse from correlated residuals $u_t$ as those always co-move. In other words, studying $u_t$ assuming everything else stays constant is generally inconsistent with the data. In sum, the Choleski decompostion is one way to transform the observed (but practically useless) $u_t$ into the very useful (but originally unobserved) fundamental shocks $\boldsymbol{\varepsilon}_t$.

The assumption underlying such an approach to orthogonalization is a causal ordering of shocks. First, it is worth cataloging the relationships, i.e., which get restricted by the ordering choice and which do not. The dynamics (lead-lag relationships as characterized by $\Psi$) of the VAR are exempted as they are already completely identified by the data itself. Rather, the ordering restricts how variables interact together \textit{within the same month}, conditional on the previous state of the system. This is done by making an explicit assumption about the composition of (reduced-form) deviations of Arctic variables from their predicted values (i.e., the anomalies). Precisely, the lower-triangular structure of \eqref{choleskimatrix} implies that residuals of a variable at position $i$ are only constituted of structural shocks $\boldsymbol{\varepsilon}_t$ of variables ordered before it. To make that explicit, we report $\boldsymbol{u}_t=C\boldsymbol{\varepsilon_t}$ in full:
\begin{align}\label{choleskimatrix}
\arraycolsep=4pt\def\arraystretch{0.557}
\begin{bmatrix}
    u_{\vphantom{1}t}^{CO_2}  \\
    u_{\vphantom{2}t}^{TCC} \\
    u_{\vphantom{3}t}^{PR} \\
   u_{\vphantom{4}t}^{AT} \\
  u_{\vphantom{5}t}^{SST} \\
  u_{\vphantom{6}t}^{SIE} \\
  u_{\vphantom{7}t}^{SIT} \\
  u_{\vphantom{8}t}^{Alb} \\
\end{bmatrix}
=
\begin{bmatrix}
    c_{11}^{} & 0_{\vphantom{11}}^{\vphantom{CO_2}}  & 0_{\vphantom{11}}^{\vphantom{CO_2}}  & 0_{\vphantom{11}}^{\vphantom{CO_2}}  & 0_{\vphantom{11}}^{\vphantom{CO_2}}  & 0_{\vphantom{11}}^{\vphantom{CO_2}}  & 0_{\vphantom{11}}^{\vphantom{CO_2}}  & 0_{\vphantom{11}}^{\vphantom{CO_2}} \\    
    c_{21}^{\vphantom{TCC}} & c_{22}^{\vphantom{TCC}}  & 0_{\vphantom{22}}^{\vphantom{TCC}}  & 0_{\vphantom{22}}^{\vphantom{TCC}}  &  0_{\vphantom{22}}^{\vphantom{TCC}} & 0_{\vphantom{22}}^{\vphantom{TCC}}  & 0_{\vphantom{22}}^{\vphantom{TCC}}  & 0_{\vphantom{22}}^{\vphantom{TCC}} \\
    c_{31}^{\vphantom{PR}}  & c_{32}^{\vphantom{PR}}  & c_{33}^{\vphantom{PR}}  & 0_{\vphantom{33}}^{\vphantom{PR}}  & 0_{\vphantom{33}}^{\vphantom{PR}}  & 0_{\vphantom{33}}^{\vphantom{PR}}  & 0_{\vphantom{33}}^{\vphantom{PR}}  & 0_{\vphantom{33}}^{\vphantom{PR}} \\ 
        c_{41}^{\vphantom{AT}}  & c_{42}^{\vphantom{AT}}  & c_{43}^{\vphantom{AT}}  & c_{44}^{\vphantom{AT}}  & 0_{\vphantom{44}}^{\vphantom{AT}}  & 0_{\vphantom{44}}^{\vphantom{AT}}  & 0_{\vphantom{44}}^{\vphantom{AT}}  & 0_{\vphantom{44}}^{\vphantom{AT}} \\      
            c_{51}^{\vphantom{SST}}  & c_{52}^{\vphantom{SST}}  & c_{53}^{\vphantom{SST}}  & c_{54}^{\vphantom{SST}}  & c_{55}^{\vphantom{SST}}  & 0_{\vphantom{55}}^{\vphantom{SST}}  & 0_{\vphantom{55}}^{\vphantom{SST}}  & 0_{\vphantom{55}}^{\vphantom{SST}} \\           
               c_{61}^{\vphantom{SIE}}  & c_{62}^{\vphantom{SIE}}  & c_{63}^{\vphantom{SIE}}  & c_{64}^{\vphantom{SIE}}  & c_{65}^{\vphantom{SIE}}  & c_{66}^{\vphantom{SIE}}  & 0_{\vphantom{66}}^{\vphantom{SIE}}  & 0_{\vphantom{66}}^{\vphantom{SIE}} \\
                    c_{71}^{\vphantom{SIT}}  &  c_{72}^{\vphantom{SIT}}  &  c_{73}^{\vphantom{SIT}}  &  c_{74}^{\vphantom{SIT}}  &  c_{75}^{\vphantom{SIT}}  &  c_{76}^{\vphantom{SIT}}  &  c_{77}^{\vphantom{SIT}}  & 0_{\vphantom{77}}^{\vphantom{SIT}} \\
   c_{81}^{\vphantom{Alb}}  & c_{82}^{\vphantom{Alb}} & c_{83}^{\vphantom{Alb}}  & c_{84}^{\vphantom{Alb}}  & c_{85}^{\vphantom{Alb}}  & c_{86}^{\vphantom{Alb}}  & c_{87}^{\vphantom{Alb}}  & c_{88}^{\vphantom{Alb}} 
\end{bmatrix}
\times
\begin{bmatrix}
    \varepsilon_{\vphantom{1}t}^{CO_2}  \\
   \varepsilon_{\vphantom{2}t}^{TCC} \\
    \varepsilon_{\vphantom{3}t}^{PR} \\
   \varepsilon_{\vphantom{4}t}^{AT} \\
 \varepsilon_{\vphantom{5}t}^{SST} \\
  \varepsilon_{\vphantom{6}t}^{SIE} \\
  \varepsilon_{\vphantom{7}t}^{SIT} \\
  \varepsilon_{\vphantom{8}t}^{Alb} \\
\end{bmatrix}.
\end{align}
Only if variable $i$ is ordered below variable $j$, will a "fundamental" shock to $j$ affect variable $i$ contemporaneously. Otherwise, variable $i$ will experience the effect of that shock with a lag of at least one month (which corresponds to one time unit in the application). For example, the $CO_2$ anomalies (which means, unpredictable by the past behavior of any of the eight variables) are assumed to be composed of structural $CO_2$ shocks only. This implies that the effect of other variables on $CO_2$ take at least a month (but perhaps more) to set in. In contrast, SIE or Albedo anomalies can be composed of a variety of fundamental shocks. Those restrictions are not without cost as the ordering of the variables may influence our understanding of the mechanism uncovered by the VARCTIC. This is why the ordering must be motivated based on the application at hand.\footnote{Moreover, when possible, the robustness of results to some reasonable alterations of the ordering should be assessed.}
\vskip 0.2cm

{\sc \noindent \textbf{Motivating The Ordering}.} It is well established that the melting SIE and the state of the Arctic environment are both results of exogenous (to other Arctic variables) human action (\cite{DaiLuoSong19}, \cite{notz2016observed}). We view the Arctic system as being subject to feedback loops that may amplify the effect of exogenous shocks way beyond their original impact. However, the original stimulus is very likely to be anthropogenic, given that without the unprecedented increase in $CO_2$ emissions and subsequent rise in global temperature, none of these mechanisms would have been so evident in effect (\cite{Ams2010}, \cite{MelRichYo14}).\footnote{ \cite{Meier2014} give an in-depth description of the various internal factors, their mutual interaction and their response to carbon dioxide.} Consequently, we order $CO_2$ first. The implication is that shocks to any of the other variables can impact $CO_2$ with a minimal delay of one month. In contrast, $CO_2$ can impact any variable in the system either contemporaneously, in the short/medium/long run, or both.


In the spirit of many medium to large BVAR applications to macroeconomic data (\cite{bernanke2005measuring}, \cite{christiano1999monetary}, \cite{stock2005implications} and \cite{banbura2010large}), we classify the variables, describing the internal climate variability, into fast-moving and slow-moving ones. TCC, PR and AT 
are classified as fast-moving. 
Absorbing short- and longwave radiation, clouds have a significant impact on the earth's energy balance and thus its overall heat content \citep{Carslaw2002}. But clouds eventually carry precipitation with not unambiguously determined effects on SIE (\cite{Park13}, \cite{Meier2014}). We order both variables before the temperature variables AT and SST.
Besides AT, also SST, especially warmer water from the Atlantic Ocean, contributed to shaping the historically unprecedented decline of SIE over the last four decades \citep{Meier2014}. Here we follow \cite{Park13} who state that besides the cooling effects of a melting ice cover, SST is highly influenced by currents and winds, transferring warmer energy from lower to higher latitudes. We therefore place SST at the boundary of fast- and slow-moving variables. 


The last block of variables comprises, SIE, SIT, and Albedo.
SIT is an underestimated determinant of how SIE reacts to both external forcing and internal variability (\cite{Meier2014}, \cite{Park13}). Thicker layers make the ice more resilient 
and increase Albedo, while thin ice is more easily advected by winds, making SIE more sensitive to extreme events
\citep{Meier2014}.
We order SIT -- and Albedo -- after SIE because we hypothesize that the effect of shocks of the former can only influence the latter with a certain delay. For instance, shocks to SIT via increased water precipitation or strong winds will immediately reduce SIT but SIE only with a certain lag. Lastly, we regard Albedo
as being driven contemporaneously by all other factors. 

To wrap up, it is worth re-emphasizing that identification, via the described ordering, is necessary to interpret and understand the mechanisms captured by the VARCTIC. However, ordering choices do \textit{not} alter forecasts. Mechanically, this happens because the potentially contentious matrix $A$ does not enter the forecasting equation \eqref{eq:it_fcst}.

{\sc \noindent \textbf{On Potentially Excluded Mechanisms}.}  We consider VARCTIC 18 in part to confirm that the key channels are already accounted for in VARCTIC 8. For example, studies have emphasized the role of incoming long- and shortwave radiation and their interactions with SIE and SIT (see \cite{BurtEtAl2016}, \cite{DaiLuoSong19}). The impact of downwelling longwave radiation (DLW) on SIE is not direct, but transmitted via DLW's influence on AT. Here, thickness of sea ice is crucial, as thinner ice is more susceptible to DLW than thicker layers \citep{ParkEtAl2015}. As we will show later (like in figure \ref{fig:18BVAR_$CO_2$_UNCOND_RCP85_RCP26_DD}), accounting for both short- and longwave radiation in VARCTIC 18, the forecast of an ice-free Arctic deviates only marginally from the ice-free date projected by VARCTIC 8. This result suggests 
that short- and longwave radiation does not impact SIE directly, but rather affects the evolution of the Arctic's sea ice cover via other variables (e.g., AT and SIT), which VARCTIC 8 already accounts for. 
In a similar line of thought, upper-ocean heat-content may also influence to the evolution of SIE. Studies have found that anomalies in the temperature of the upper-ocean layers and anomalies in SST do coincide \citep{ParkEtAl2015}, making an extension of both VARCTIC models dispensable. 

However, it is not excluded that some non-local processes (e.g., poleward atmospheric heat transport) do contribute to sea-ice loss through channels not represented in both VARCTICs. As stated earlier, we opted for including local processes only (in addition to $CO_2$) because this makes the VARCTIC a complete system where all $M$ variables are internally modeled and forecasted jointly. Adding non-local processes raises the additional question of how to model their external dependence, a complication left for future research.\footnote{The literature on Global VARs could provide a natural place to start \citep{pesaran2009forecasting}.}

\subsection{Impulse Response Functions}\label{sec:irf}

Since \cite{sims1980}, the dominant approach for studying the properties of the VAR around its deterministic path has been impulse response functions (IRFs) to structural shocks. Thanks to the orthogonalization strategy discussed in \ref{sec:ID}, we converted plain regression residuals into orthogonal shocks.\footnote{Mathematically, we took a linear combination of the VAR residuals (an unpredictable change in a variable of interest, $\boldsymbol{u}_t$) such that $\boldsymbol{u}_t=C\boldsymbol{\varepsilon_t}$.} The dynamic effect of these specific disturbances (the impulse) can be analyzed as that of a randomly assigned treatment.\footnote{Of course, one could look at how the system responds to an impulse from a residual $u_{m,t}$, but the interpretation will be rather weak because those are correlated across equations.} Uncorrelatedness of $\varepsilon_{m,t}$ implies the "keeping everything else constant" interpretation -- hence, a causal meaning for IRFs -- is guaranteed by construction. 

It is natural to wonder about the meaning of uncorrelated shocks in a physical system. Mechanically, these shocks are the difference between the realized state of a variable and its predicted value as per the previous state of the dynamic system. These unpredictable anomalies, which emerge from outside a well-specified VARCTIC, are the key to understanding the dynamic properties of the model. A now obvious example of a shock will be that of $CO_2$ emissions reduction in 2020: it is inevitable that the observed emissions will be lower than what was predicted by the endogenous system since the latter excludes "pandemics". Any model that is partially incomplete will be subject to external shocks. The study of such exogenous impulses may be alien-sounding, especially when contrasted with the deterministic environment of a climate model. Nonetheless, understanding the properties of a climate model by conditioning on a particular RCP scenario is equivalent to conditioning on a series of shocks. Hence, one can understand the VARCTIC and its IRFs as expanding the number of potentially exogenous sources of forcing. Of course, our later focus on $CO_2$ shocks is expressively motivated by the fact that the latter is a well-accepted source of exogenous forcing in climate systems.

The impulse response function of a variable $m$ to a one standard deviation shock of $\varepsilon_{\tilde{m},t}$ is defined as
\begin{align}
IRF(\tilde{m}\rightarrow m, h) =E(y_{m,t+h}|\boldsymbol{y_{t}},\varepsilon_{t,\tilde{m}}=\sigma_{\varepsilon_{\tilde{m}}})-E(y_{m,t+h}|\boldsymbol{y_{t}},\varepsilon_{t,\tilde{m}}=0).
\end{align}
Thus, it is the expected difference, $h$ months after "impact", between an Arctic system that responded to an unexpected $CO_2$ increase, and the same system where no such increase occurred. In a linear VAR with one lag ($P=1$), the IRF of \textit{all variables} can easily be computed from the original estimates using the formula
\begin{align}
IRF(\tilde{m}\rightarrow \boldsymbol{m}, h) = \Psi^h A^{-1} e_{\tilde{m}}
\end{align}
where $e_{\tilde{m}}$ is a vector with $\sigma_{\varepsilon_{\tilde{m}}}$ in position $\tilde{m}$ and zero elsewhere. This just means that we are looking at the individual effect of $\varepsilon_{\tilde{m}}$ while all other structural disturbances are shut down.\footnote{In the case of a linear VAR with $P>1$ lags, we must use the companion matrix form. \textcolor{black}{The relevant formula (equation \eqref{IRF_VARP}) can be found in the discussion of appendix \ref{sec:TMAtheory}}.
}

The latter discussion focused on analyzing how our dynamic system responds to an external/unforeseeable impulse, which is a standard way of interpreting VAR systems. Of course, we are also interested in the "systematic" part of the VAR that is responsible for the propagation of shocks when they do occur -- the IRF transmission mechanism. In section \ref{sec:sensitivity}, we focus our attention on $CO_2$ and AT shocks and quantify the amplification effect of different channels.

\subsection{Bayesian Estimation}\label{sec:bayes}

We use a Bayesian VAR in the tradition of \cite{litterman1980}. There are two crucial advantages of doing so. First, Bayesian inference does not depend on whether the VAR system is stationary or not (\cite{FanWen92}). We are effectively modeling variables in levels and expecting at least one \textit{explosive} root. \textit{Frequentist} inference is notoriously complicated in such setups (\cite{almosteverythingaboutUR}) and even standard approaches for non-stationary data have well-known robustness problems (\cite{elliott1998}). From a practical point of view, using non-stationary data means that standard test statistics (like Granger Causality tests) will be undermined by faulty standard errors, potentially leading to erroneous conclusions. 


Second, for us to consider a system of many variables estimated with a relatively small number of observations, Bayesian shrinkage can be beneficial to out-of-sample forecasting performance and help in reducing estimation uncertainty (like those of IRFs). In fact, VARs are known to suffer from the curse of dimensionality as the number of parameters scales up very fast with the number of endogenous variables.\footnote{Such a situation motivates \cite{McGraw19} to use the LASSO.} Via informative priors, Bayesian inference provides a natural way to impose soft/stochastic constraints (that is, constraints are not imposed to bind) and yet keep inference possible \citep{banbura2010large}.\footnote{For instance, running a VAR with LASSO would induce some form of shrinkage but inference is far from easy.} Furthermore, we are interested in transformations (forecasting paths, impulse response functions) of the parameters rather than the parameters themselves. Inference for such objects can easily be obtained by transforming draws from the posterior distribution. All these procedures are well established in the macroeconometrics community and packages are available in most statistical programming software \citep{BearMatlab}. An extended discussion of the prior, its motivation for time series data and details on the exact values of (data-driven) hyperparameters used, can all be found in section \ref{sec:priors}. 

Finally, the maximal lag order of the VAR, $P$ in equations \eqref{rf_VAR} and \eqref{struct_VAR}, must be chosen.\footnote{To re-emphasize, this means $y_{t-p}$ for  $p=1,...,P$ are included.}  Its selection is yet another incarnation of the bias-variance trade-off. We fix the number of lags in VARCTIC 8 to $P=12$ and to $P=3$ in VARCTIC 18 respectively. That choice is based on the Deviance Information Criterion (DIC) -- the Bayesian analog to popular information criteria used for model selection. Accordingly, the superior VARCTIC 8 would set $P=3$ (DIC=-6988)\footnote{The lower, the better.}, a choice which only provides a marginal improvement with respect to $P=12$ (DIC=-6894).\footnote{Additionally, the reported DIC for $P=12$ is superior to other natural candidates such as $P=1$ and $P=24$.} Since structural analysis is an essential part of this paper, we err on the side of having slightly higher variance, but potentially richer dynamics for IRFs. In the large VARCTIC 18, the need for shrinkage is magnified and $P=3$ is the obvious more reasonable choice.

An extraneous question, which can benefit from verification by DIC, is whether trends should be included. We hypothesize that VARCTIC 8 is a complete, divergent system which can endogenously explain the trending behavior of all its variables by the joint action of $CO_2$ forcing and feedback loops. If that were not to be true, including linear trends would noticeably improve model fit, and lower the DIC even further. Backing our claim that the VARCTIC needs no additional (and hardly climatically-explainable) statistical crutch, the DIC from including trends is worse (now DIC=-6817 for VARCTIC 8) than that of the original model.

\section{Results} \label{sec:results}

A VAR contains many coefficients -- there are $8 \times (8 \times 12 +1)= 776$ in the baseline VARCTIC.\footnote{The same arithmetic gives a total of 990 parameters in VARCTIC 18.} Staring at them directly is unproductive and a single coefficient (or even a specific block) carries little meaning by itself. As it is common with VARs in macroeconomics, we rather study the properties of the VARCTIC by looking at its implied forecasts and its IRFs.

\subsection{The "Business as Usual" Forecast}
\begin{figure}[h!]
\centering
\includegraphics[scale=0.35]{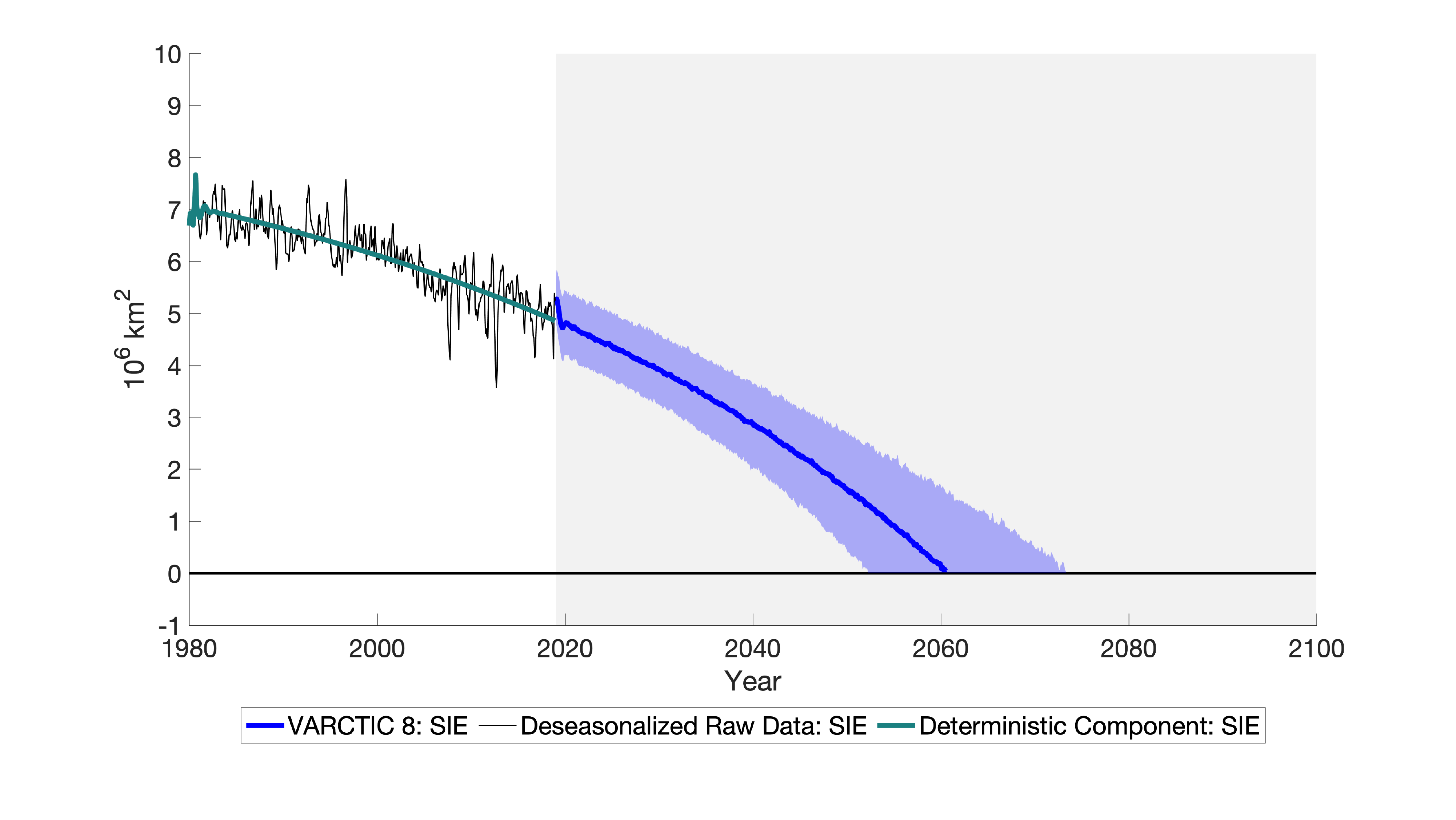}
\vspace*{-0.5cm}
\caption{Trend Sea Ice Extent for September. \\ Shade is the 90\% credible region.}\label{fig:SIE_8dim}
\end{figure}

We report here the unconditional forecast of our main VARs. VARCTIC 8 suggests SIE to hit the zero lower bound around 2060 (see Figure \ref{fig:SIE_8dim}), whereas VARCTIC 18 projects the Arctic to be ice-free at about the same time (see Figure \ref{fig:18BVAR_$CO_2$_UNCOND_RCP85_RCP26_DD}).\footnote{We include in the graph the in-sample deterministic component of the VAR (as discussed in \cite{PLR}, which is essentially a long-run forecast, starting from 1980 (the same sort of which we are doing right now for the next decades) using the VAR estimates of 12 lags.} 
The shaded area shows 90\% of all the potential paths of the respective VARCTIC. That is, VARCTIC 8 dates the Arctic to be totally ice-free for the first time somewhere between 2052 and 2073 with a probability of 90\%. VARCTIC 18 slightly extends that time frame to the year 2079.

For the two models, the median scenario has SIE being less than 1 $times$ $10^6$ km$^2$ by 2054 and 2060 respectively. The 1 $times$ $10^6$ km$^2$ is more likely an interesting quantity since the "regions north of Greenland/Canada will retain some sea ice in the future even though the Arctic can be considered as 'nearly sea ice free' at the end of summer." (\cite{wang2009sea}). The corresponding credible regions mark the period 2047-2065 for VARCTIC 8 and 2047-2069 for VARCTIC 18 respectively. These dates and time spans range in the close neighborhood of previous climate model simulations (see \cite{JahnEtAl2016}). For both VARCTICs, less than 5\% of the simulated paths hit 0 before 2050, making it an unlikely scenario according to our calculations. 
In essence, the two models suggest SIE melting at a rate that is slower than \cite{DieRu19}'s results, but much faster than most CMIP5 models \citep{StroeveEtAl12}, and in line with the latest CMIP6 calculations \citep{Notzetal2020}.\footnote{Note that augmenting VARCTIC 8 with other greenhouse gases such as methane (CH$_4$) procures near-identical results (i.e., forecasting and forthcoming IRFs). This reinforces the view that $CO_2$ plays a distinct and important role in determining the fate of SIE.} 

Nonetheless, it is natural to ask how much we can trust a forecast made 40 years ahead, based on 40 years of data behind. To a large extent, answering this amounts to catalog what types of uncertainty the 90\% credible regions incorporate, and those they do not. These reflect both forecasting uncertainty (the presence of shocks) and parameter uncertainty. The latter means the 90\% regions reflect what happens to the spread of forecast paths when small disturbances are incorporated in the (estimated) coefficient matrix. In other words, those bands conveniently (and correctly) quantify prediction uncertainty accounting for the fact that we are iterating something that is estimated. All things considered, uncertainty is correctly calibrated as long as the model is correctly specified. As it is the case with any statistical approach, the VARCTIC necessarily assumes that the physical reactions estimated on previous decades' data remain valid for those to come. Thus, if the future holds unprecedented nonlinear mechanisms or previously undetectable relationships\footnote{\cite{notz2016observed} find that in nearly all CMIP5 models the negative relationship between $CO_2$ and SIE was not prevalent until the second half of the twentieth century.}, 
the VARCTIC can hardly accommodate for that. In contrast, any intensification of phenomena characterized by our 8 key variables (like Albedo feedback) should be successfully captured out-of-sample. With VARCTIC results being in accord with the recent CMIP6 consensus, our specification seems to capture the main drivers and dynamics of the Arctic ecosystem.\footnote{Though, we acknowledge recent research, which stresses the role of ozone depleting substances (ODS) -- another form of anthropogenic greenhouse gases -- in the warming of the Arctic region over the last decades \citep{PolvaniEtAl2020}.} Finally, future $CO_2$ emissions are an uncontested source of uncertainty for long-run SIE forecasts. Section \ref{policy} studies how those (and their credible regions) behave under standard forcing scenarios.

\subsection{Impulse Response Functions}

\begin{figure}[h!]
\centering
\includegraphics[width = \textwidth]{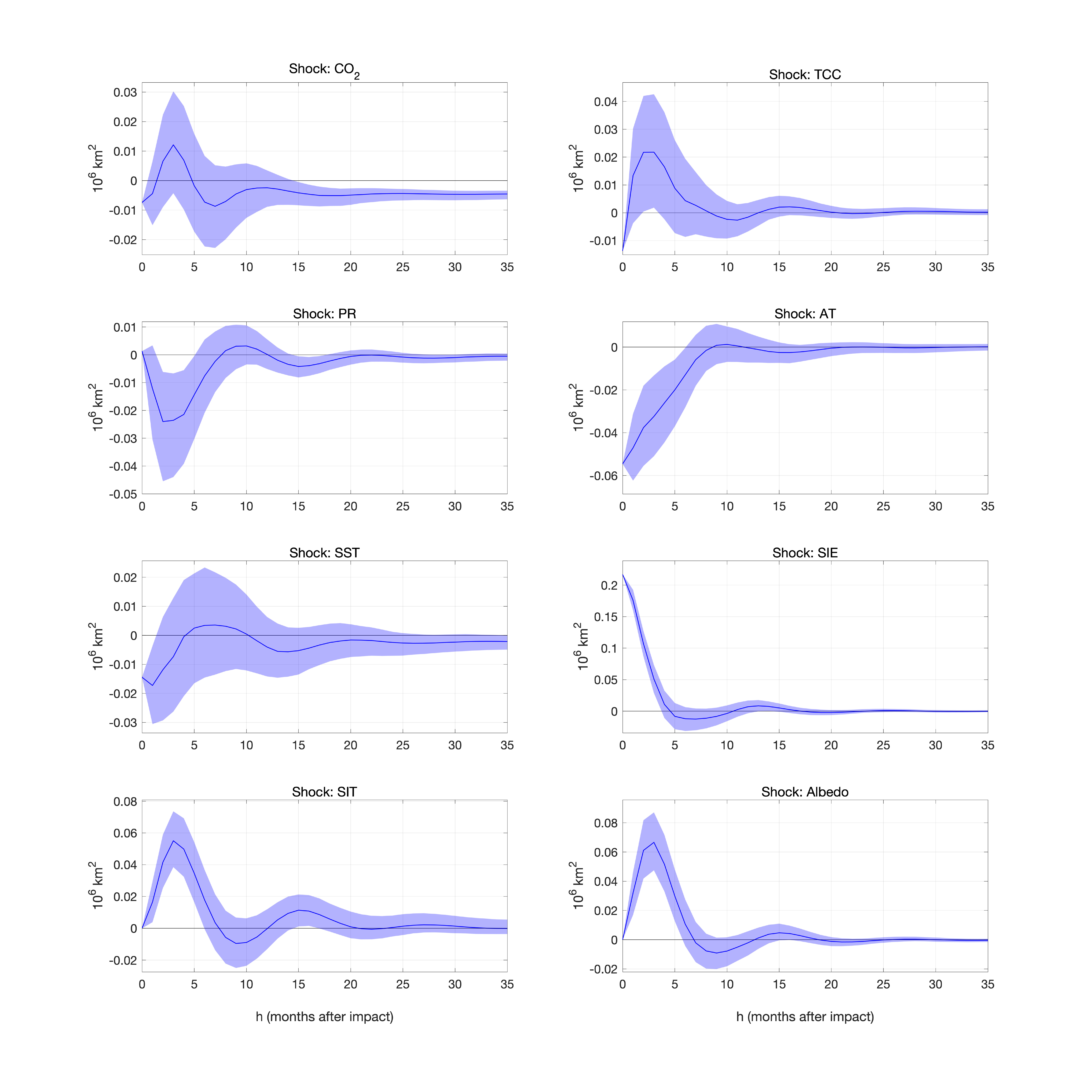}
\vspace*{-1cm}
\caption{IRFs: Response of Sea Ice Extent. \\ Shade is the 90\% credible region.}\label{fig:SIE_8dim_IRF}
\end{figure}

Figure \ref{fig:SIE_8dim_IRF} displays impulse response functions -- the response of SIE to a \textit{positive} shock of one standard deviation to any of the model's \emph{M} variables. To reflect parameter uncertainty, we additionally report the 90\% credible region for each IRF. This means the gray bands contain 90\% of the posterior draws from VARCTIC 8. Those are crucial to determine whether the attached IRFs describe significant physical phenomena or not. Particularly, when the credible region extends to both positive and negative sides, the IRF characterizes a phenomenon of negligible importance. In those instances (e.g., many IRFs at horizon $h>24$ months), the posterior mean's (black line) difference from 0 could merely be due to parameter uncertainty, and can be thought of as approximately 0. 

\begin{figure}[t]
\centering
\includegraphics[width=1\textwidth, angle=0]{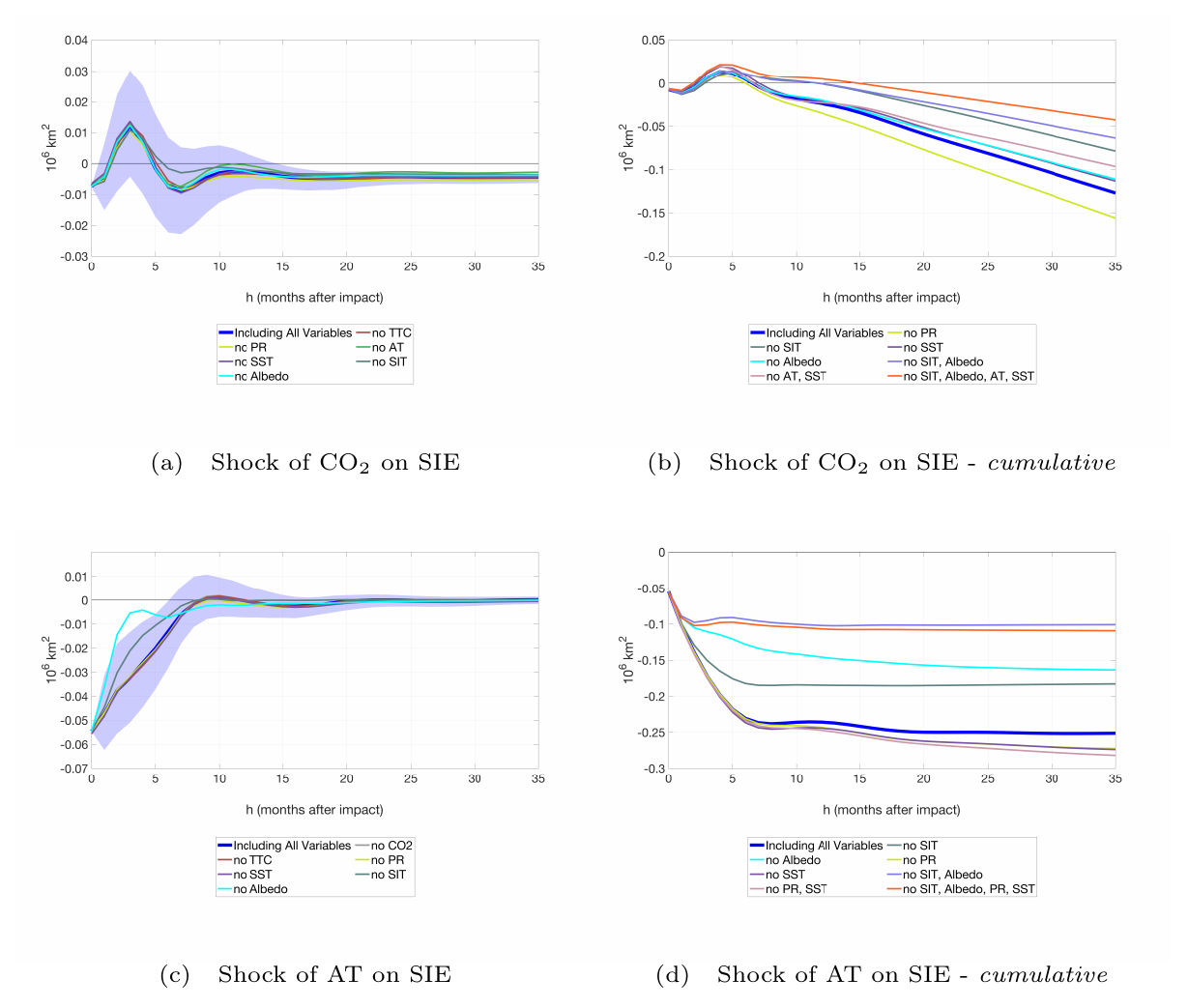}\\
\caption{IRF Decomposition: Response of Sea Ice Extent. \\ Shade is the 90\% credible region for the original IRF.}\label{fig:IRF_TMA_dim8_CO2AT}
\end{figure}

The resulting impact of $CO_2$ anomalies on SIE is sizable and most importantly, durable. While the sign of the response is highly uncertain and weak for more than a year, $CO_2$ shocks emerge to have a lasting downward effect on SIE. The relevance of the $CO_2$/SIE relation is not a surprise (\cite{notz2016observed}). Moreover, this behavior is distinct from other shocks that rather have a significant short-run effect but no significant effect after more than roughly six months. More precisely, the effect of $CO_2$ impulses more than a year to settle in (not significant for approximately 15 months) but ends up having a continuing downward effect on trend SIE of approximately -0.005 $10^6$ km$^2$. This mechanically implies that a one-off $CO_2$ deviation from its predicted value/trend leads to a \textit{cumulative} impact that is ever increasing in absolute terms (as displayed later in Figure \ref{fig:IRF_TMA_dim8_CO2AT}(b)). It is important to remember that this is the effect of an \textit{unexpected} increase in $CO_2$ which is to be contrasted with the systematic effect that will be studied later. However, in the framework of this section -- where $CO_2$ is allowed to endogenously respond to Arctic variables -- this is as close as one can get to obtain an experimental/exogenous variation needed to evaluate a dynamic causal effect. -0.005 $10^6$ is roughly 0.1\% of the last deterministic trend value of SIE. $CO_2$ shocks, by construction of our linear VAR, have mean 0 and there are approximately as many positive and negative shocks in-sample. The linearity and symmetry of the VAR imply that these durable effects are present for both upward and downward deviations from the deterministic trend. 

Other shocks have sizeable impacts that eventually vanish, which is the traditional IRF shape one would expect to see from a VAR on macroeconomic data. For instance, AT and Albedo IRFs clearly have the expected sign. However, they do not have the striking lasting impact of $CO_2$ perturbations. To rationalize the short-lived $IRF(\text{AT}\rightarrow \text{SIE})$, it is worth re-emphasizing what is meant by an AT shock. It is an AT anomaly that is not explicable by (i) the previous state of the system and (ii) other structural shocks ordered before it ($CO_2$, TTC, PR). As an example, one could think of the 2007 record low SIE (at that time) being attributed to an abnormally high “atmospheric flow of warm and humid air” from lower latitudes into the Arctic region \citep{graversen2011}. As we will see in section \ref{sec:sensitivity}, a $CO_2$ shock triggers (with a significant delay) a persistent increase in AT, which eventually impacts SIE downward through the systematic part of the VAR. Thus, the short-lived response of SIE to AT shocks does not rule out a lasting impact of AT on SIE. Rather, it means that when it occurs, the origin of the anomaly is not AT itself, but likely $CO_2$.

Similar to an unforeseeable AT shock, a one-time Albedo shock will \textit{not} have a lasting effect on SIE neither. 
This does not preclude Albedo to amplify other shocks as we will see in the next section.\footnote{For a discussion of VARCTIC 18 results, see section \ref{sec:v18}.} 
Finally, a rightful concern is  whether IRFs remain unaltered upon sensibly altering the ordering of section \ref{sec:ID}. To a large extent, they do. For instance, placing SST and AT before TCC and PR brings no noticeable change. So does moving Albedo from last to second \textcolor{black}{(see section \ref{sec:altord})}.

\subsection{Amplification of $CO_2$ and Temperature Shocks by Feedback Loops}\label{sec:sensitivity}

The melting of SIE is happening much faster than many other phenomena that are also believed to be set in motion by the steady increase of $CO_2$ emissions. Many recent papers (\cite{notz2012observations}, \cite{wang2012sea}, \cite{serreze2015arctic}, \cite{notz2017arctic}) argue with theory/climate models or correlations that external $CO_2$ forcing is responsible for the long-run trajectory of SIE. Some of these 
findings led  \cite{notz2016observed} to conclude that climate models severely underestimate the impact of $CO_2$ on SIE.

A rather consensual view is that the very nature of the Arctic system leads to the \textit{amplification} of such external forcing shocks. 
An understanding -- from observational data -- on how the Arctic may amplify -- or not -- certain external forces is still pending. Fortunately, a VAR can quantify the contribution of different variables in explaining how a dynamic system responds to an external impulse.

\subsubsection{Methodology}
A potential approach that has a long history in econometrics is the use of Granger Causality (GC) tests. Those consist of evaluating predictive causal statements (such as $X_{t-1} \rightarrow Y_{t}$ and $Y_{t-1} \rightarrow X_{t}$) via significance tests in time series regressions. They have been recently advocated for climate applications by \cite{mcgrawbarnes2018}. Nevertheless, those tests often fall short of answering questions of interest. First, the meaning of the test is not obvious when more than two variables are included and/or if one is interested in multi-horizon impacts. Second, in the wake of a GC test rejection, the block of reduced-form coefficients\footnote{Precisely, we mean coefficients on lags of $X_t$ in a regression of those on $Y_t$ (including lags of $Y_t$ as well).}, which we know to be of some statistical importance, are very hard to interpret. In other words, we know some channel matters, but we have little idea \textit{how} it matters.\footnote{Similar concerns led us to discard \cite{liang2014}'s quantitative causality since the currently available formula only applies to bivariate systems. Further, it does not allow for contemporaneous relationships which are clearly present in our application (and a feature of almost any discretely sampled multivariate time series).} 

In light of the above, we rather opt for IRF \textit{Decomposition}. As the name suggests, the physical reaction characterized by IRFs will be decomposed as a sum of transmission channels, which contributions' magnitudes and signs are directly informative. Less abstractly, the consequential negative response of SIE to $CO_2$ shocks is likely composed of a direct effect and many entangled indirect effects (e.g., that of AT and Albedo). Understanding those in the dynamic setup of a VAR is much more intricate than in a static regression setting. This is so because IRFs -- for horizons greater than one -- are obtained by iterating predictions, which means $X$ can impact $Y$ through $Z$, but also through any of its lags. We employ a strategy that has been used in macroeconomics to better understand the transmission mechanism in VARs. It consists, rather simply, of shutting down "channels" and plotting what the response to a shock would be, given that this very channel had been shut (\cite{sims2006does}, \cite{bernanke1997systematic},  \cite{bachmann2012confidence}). We can deploy this methodology to find and quantify the most important channels through which $CO_2$ and temperature shocks impact SIE.

\subsubsection{Amplification of $CO_2$ Shocks}

For VARCTIC 8, figures \ref{fig:IRF_TMA_dim8_CO2AT}(a) and \ref{fig:IRF_TMA_dim8_CO2AT}(b) show the responses of SIE to an unexpected increase in one standard-deviation of $CO_2$. The \emph{blue} line pictures the case of the \emph{baseline} VARCTIC 8 with 90\% credible region. The remaining six lines show the response of SIE to the same shock but shutting down key transmission channels. In terms of implementation, it consists of imposing \emph{hypothetical} shocks to one of the other variables which off-sets \emph{their} own response to a $CO_2$ shock.\footnote{See \cite{bachmann2012confidence} for details.} 

{The top panel of figure \ref{fig:IRF_TMA_dim8_CO2AT}} reveals -- without great surprise -- the importance of temperature (especially AT) in translating $CO_2$ anomalies into decreasing SIE. That is, we observe that shutting down these channels leads to a smaller absolute response which means that those variables can be considered as \textit{amplification channels}. Given the atypical shape of the $CO_2$ IRF, the scale of figure \ref{fig:IRF_TMA_dim8_CO2AT}(a) makes less visible the action of channels that only alter the longer-run effect. Since those effects are durably negative (at different levels), their cumulative effect will more clearly reveal their relative importance. Thus, figure \ref{fig:IRF_TMA_dim8_CO2AT}(b) displays the cumulative impact of selected (more important) channels. The two temperature channels are responsible for approximately one fourth of the cumulative effect of $CO_2$ on SIE after 3 years. More precisely, restricting temperature variables to \textit{not} respond to a positive $CO_2$ shock, decreases (in absolute terms) the after-3-years impact from -0.13 $10^6$ km$^2$ to -0.1 $10^6$ km$^2$. Of course, it was expected that temperature should be a major conductor of such shocks. We also observe similar quantitative effects for both SIT and Albedo in isolation. Most strikingly, we find that the conjunction of the Albedo and SIT amplification channels is responsible for amplifying the effect of $CO_2$ shocks by a non-negligible 50\%.

The Albedo-amplification matches evidence reported in several studies (see \cite{Pero2012}, \cite{Bjoerk2013}, \cite{Park13}) using various different methodologies. In contrast, our results for SIT-amplification contribute to a literature where a consensus has yet to emerge. The ice-growth-SIT feedback describes the observation that a thinning of the sea ice cover induces more rapid ice formation during winter, a compensating effect which \textit{slows down} melting \citep{bitz2004mechanism,GoosseEtAl2018}. Other studies have emphasized the \emph{positive feedback} between $SIT$ and $SIE$, where a thinning ice cover further accelerates melting by being less resilient to climate forcing \citep{Park13,Kwok2018}. Our results unequivocally support the latter to be most empirically prevalent.\footnote{It is plausible that the ice-growth-SIT feedback explains why both $IRF(\text{SIT}\rightarrow \text{SIE})$ (figure \ref{fig:SIE_8dim_IRF}) and SIT's influence on $IRF(CO_2\rightarrow \text{SIE})$ (figure \ref{fig:IRF_TMA_dim8_CO2AT}) take over 6 months to completely settle in --- its seasonal character dampens the (early) positive feedback effect.}
Nearly identical results are obtained when using AT, Albedo, PR, and TCC averaged between 60$^{\circ}$N and 90$^{\circ}$N latitude, suggesting most (if not all) of the action comes from higher latitudes -- hence our focus on local processes when explaining those results.


This section focused on how and why SIE responds to $CO_2$ \textit{shocks}. In section \ref{policy}, we rather look at the effect of the \textit{systematic} increase of $CO_2$ level. 

\subsubsection{Amplification of Air Temperature Shocks}
\textit{AT-shocks} are movements in AT that are not predictable given the past state of the system and are orthogonal to other shocks in the system, most notably $CO_2$. In other words, we are looking at the effect of unexpected higher/lower AT that is uncorrelated with other shocks in the system. As we saw in figure \ref{fig:SIE_8dim_IRF}, such AT anomalies have a pronounced short-run effect on trend SIE for no longer than ten months after the shocks. This means that unlike $CO_2$, the cumulative effect of AT disturbances stabilizes about 1.5 years after the event. 

{In figure \ref{fig:IRF_TMA_dim8_CO2AT}(c)}, we clearly observe (again) an important role for the thinning of ice and the Albedo effect amplifying the response of SIE to AT shocks. In fact, we see in figure {\ref{fig:IRF_TMA_dim8_CO2AT}(d)} that without them, the long-run impact is the same as the instantaneous one. Thus, this is evidence to suggest that the AT shock's long-run cumulative impact of -0.24 $10^6$ km$^2$ is mostly a result of the action of feedback loops. 

\subsection{Forecasting SIE Conditional on $CO_2$ Emissions Scenarios} \label{policy}

If $CO_2$'s trend is mostly or solely affected by factors outside of those considered in the VAR, the forecast of SIE can be improved by treating $CO_2$ forcing as exogenous and using an external forecast rather than the one internally generated by the VAR. Conditional forecasting can be achieved in VARs following the approach of \cite{waggoner1999conditional}. As we will see, this will markedly sharpen the bands around our forecasts, suggesting that a great amount of uncertainty is related to the future path of $CO_2$ emissions. Additionally, this brings the VARCTIC conceptually closer to standard analyses on the future of the Arctic (\cite{StroeveEtAl12}, \cite{stroeve2018}, \cite{Notzetal2020}).

\begin{figure}[h!]
\centering
\begin{subfigure}[h]{1\textwidth}
\centering
\includegraphics[scale=0.35]{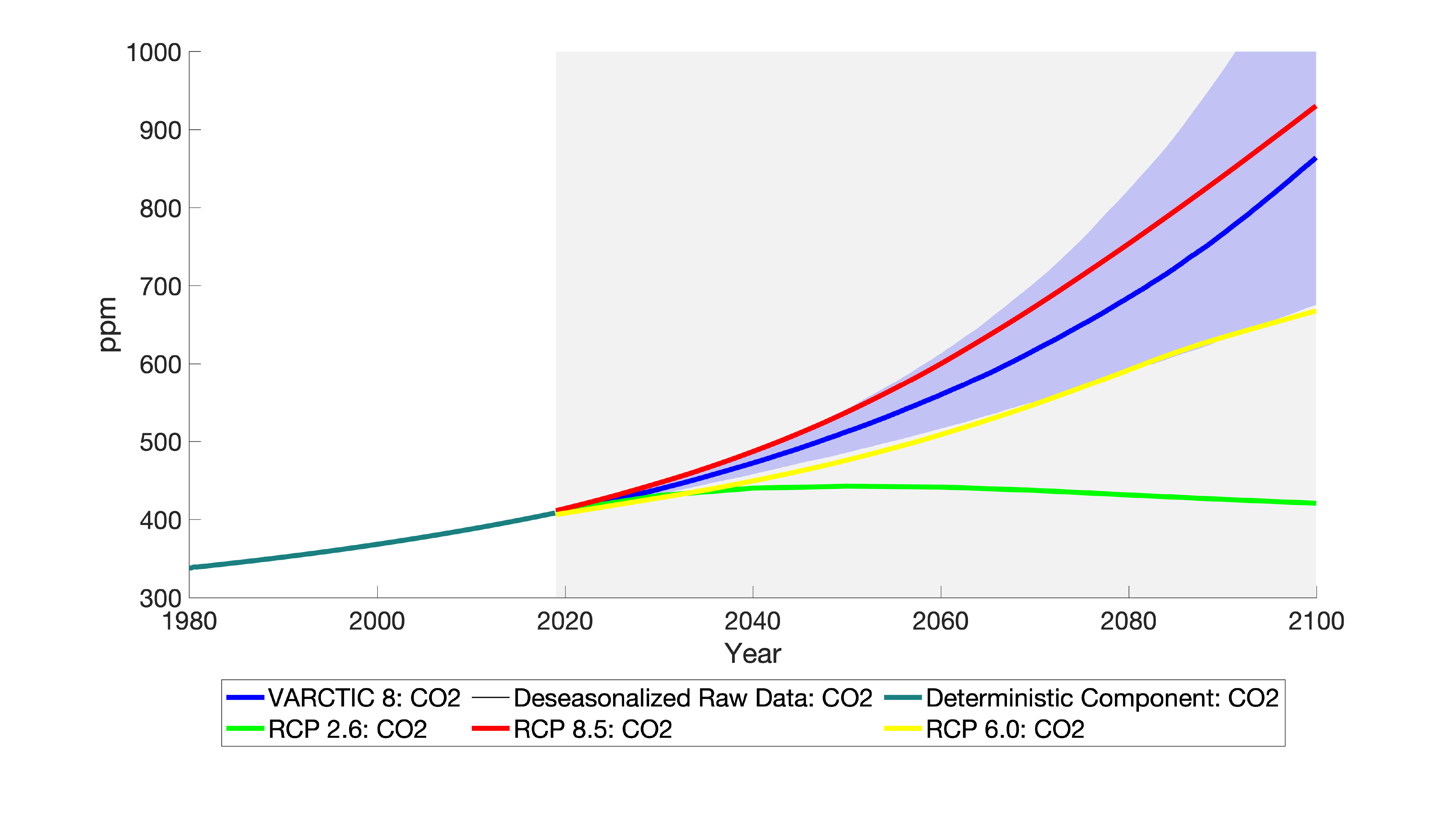}
\caption{Evolution of CO$_2$ emissions until the End of the Century under different Scenarios} \label{fig:8BVAR_Projections_DiffRCPs_CO2}
\end{subfigure}
\vskip 5mm
\begin{subfigure}[h]{1\textwidth}
\centering 
\includegraphics[scale=0.35]{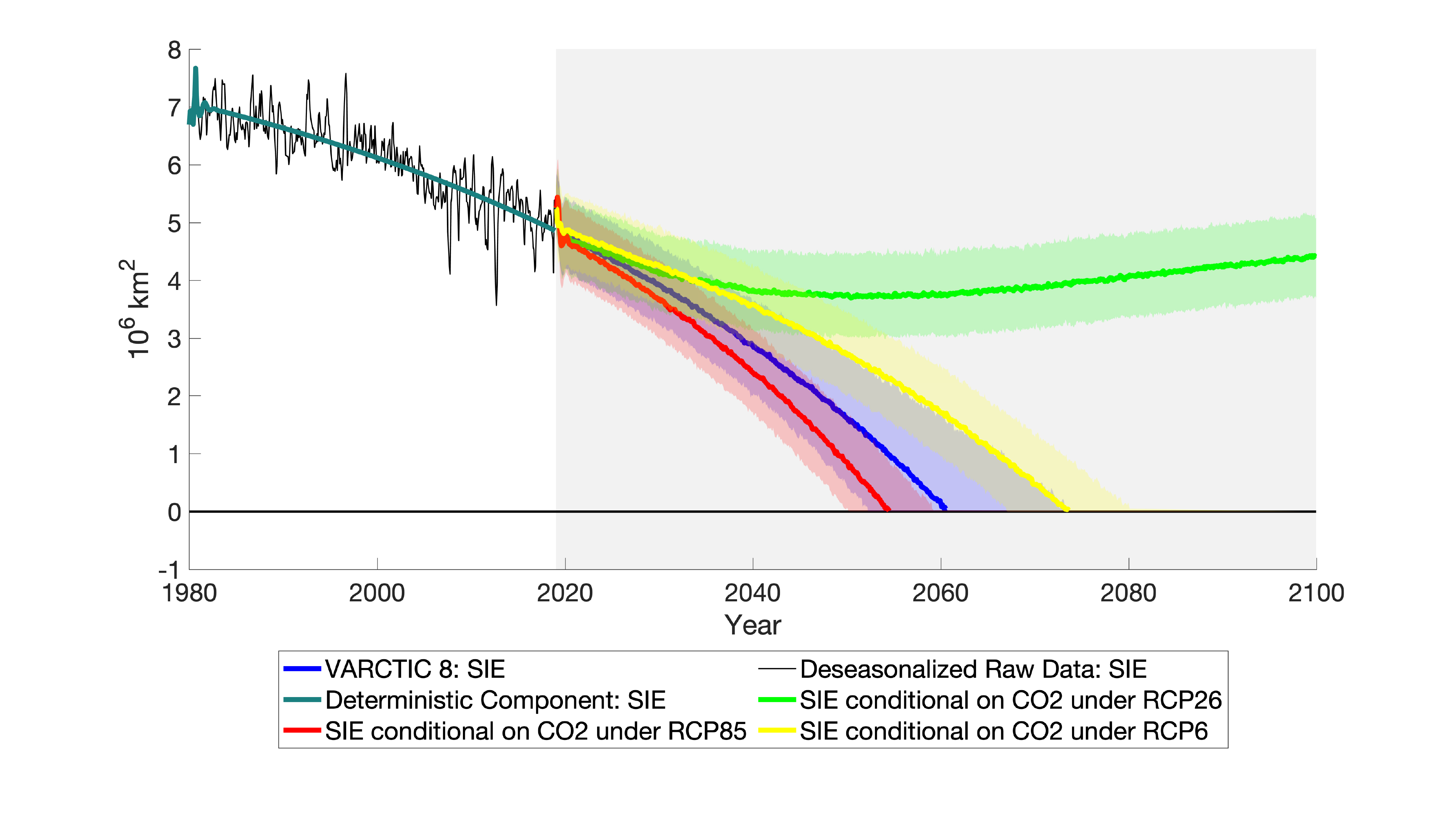}
\caption{Evolution of SIE under different Scenarios of $CO_2$} \label{fig:8BVAR_Projections_DiffRCPs_SIE}
\end{subfigure}
\caption{VARCTIC Projections \& Different RCPs. \\ Shade is 90\% credible region.} \label{fig:8BVAR_Projections_DiffRCPs}
\end{figure}

{In the spirit of \cite{SigmondEtAl2018}, who constrain the levels of AT in their climate model, we will look at $CO_2$ emissions under three different representative concentration pathways (RCP) and investigate their impact on the evolution of Arctic sea ice.}  
Figure \ref{fig:20Vars_ds} shows a steady increase in $CO_2$ emissions over the last three decades, but several RCPs paint different pictures for the trajectory of carbon emissions until the end of the century.
Figure \ref{fig:8BVAR_Projections_DiffRCPs_CO2} shows the different paths of $CO_2$ under \emph{RCP 2.6}, \emph{RCP 6}, and \emph{RCP 8.5}, as well as the projected path following VARCTIC 8. Most interestingly, we find our completely endogenous and unconditional forecast of $CO_2$ to lay somewhere between the "very bad" \emph{RCP 8.5} scenario and the "business-as-usual" \emph{RCP 6} one.

Figure \ref{fig:8BVAR_Projections_DiffRCPs_SIE} shows VARCTIC 8's projection of Arctic SIE when conditioning the out-of-sample path of $CO_2$ on the three different RCP scenarios. For reference, the figure also includes projected SIE with the future path of $CO_2$ endogenously determined within the model, as discussed above. The pictured effect is dramatic: if emissions were reduced as to follow the \emph{RCP 2.6} scenario, whose $CO_2$ emissions are still at the higher boundary of what the Paris Agreement demands, the Arctic would be far from \emph{blue} and even recover earlier losses by the end of the century. If emissions follow the more likely \emph{RCP 6}, SIE would vanish later than projected by the unconditional VARCTIC 8, but would still be completely gone during the 2070's. In the worst-case scenario, \emph{RCP 8.5}, we obtain an ice-free September by the mid-2050's. Interestingly, this result is very close to what \cite{stroeve2018} reported using a very different methodology (extrapolating a linear relationship). Their bivariate (SIE and $CO_2$) analysis suggests the Arctic summer months to be ice-free by 2050. However, in contrast, our results are much more optimistic than theirs in terms of SIE conditional on the (rather unlikely) \emph{RCP 2.6} scenario. Such analysis is not conditional on the identification scheme since it is based solely on the reduced form.\footnote{Important to note is the fact that the very last in-sample observations for $CO_2$ even range above the \emph{RCP 8.5} values, which generates the slight upward jump in case of the latter scenario.} Overall, these results reinforce the view that anthropogenic $CO_2$ is the main driver behind the current melting of SIE as well as the main source of uncertainty around the future SIE path. Furthermore, the optimistic \emph{RCP 2.6} results suggest that internal variability by itself cannot lead to the complete melting of SIE, even when starting from today's level. Overall, the VARCTIC yields similar conclusions about the importance of $CO_2$ to that of \cite{DaiLuoSong19} and \cite{notz2016observed}. It is reassuring to see that climate models' conclusions can be corroborated by a transparent approach that relies solely and directly on the multivariate time series properties of the observational record. 

\subsection{Amplification Effects in the Projection of SIE under different RCPs} \label{sec:TMA_Forecast}
The previous section documented the evolution of SIE conditional on several $CO_2$ trajectories, treating the latter as an exogenous driver. This section seeks to quantify the importance of feedback effects when it comes to translating an RCP path into SIE loss. That is, we attempt to quantify to which extent the Albedo- and SIT-effects can be held responsible for amplifying the impact of $CO_2$ forcing and thus accelerating the melting of SIE.

Following the findings of section \ref{sec:sensitivity}, in which we identified SIT and Albedo to carry potential for mitigating the adverse influence of $CO_2$ on SIE, we ask the question about how SIE would evolve, if SIT and Albedo were to remain constant at a certain level over the forecasting period. In particular, we repeat the forecasting exercise of the previous section for 
all three RCP scenarios, but keep SIT and Albedo constant until the end of the forecasting period. For both variables we set the level equal to the value, which is given by the series' deterministic component at the end of the sample period. By doing so, we create artificial shocks to both SIT and Albedo in each forecasting step, which off-set their response to the external forcing variable. 
As we are modeling a dynamic and interconnected system, these shocks do affect all the other variables (except for $CO_2$ on which we condition our forecast). 

\begin{figure}[h!]
\centering
\vspace*{-1cm}
\begin{subfigure}{\textwidth}
\centering
\includegraphics[scale = 0.35]{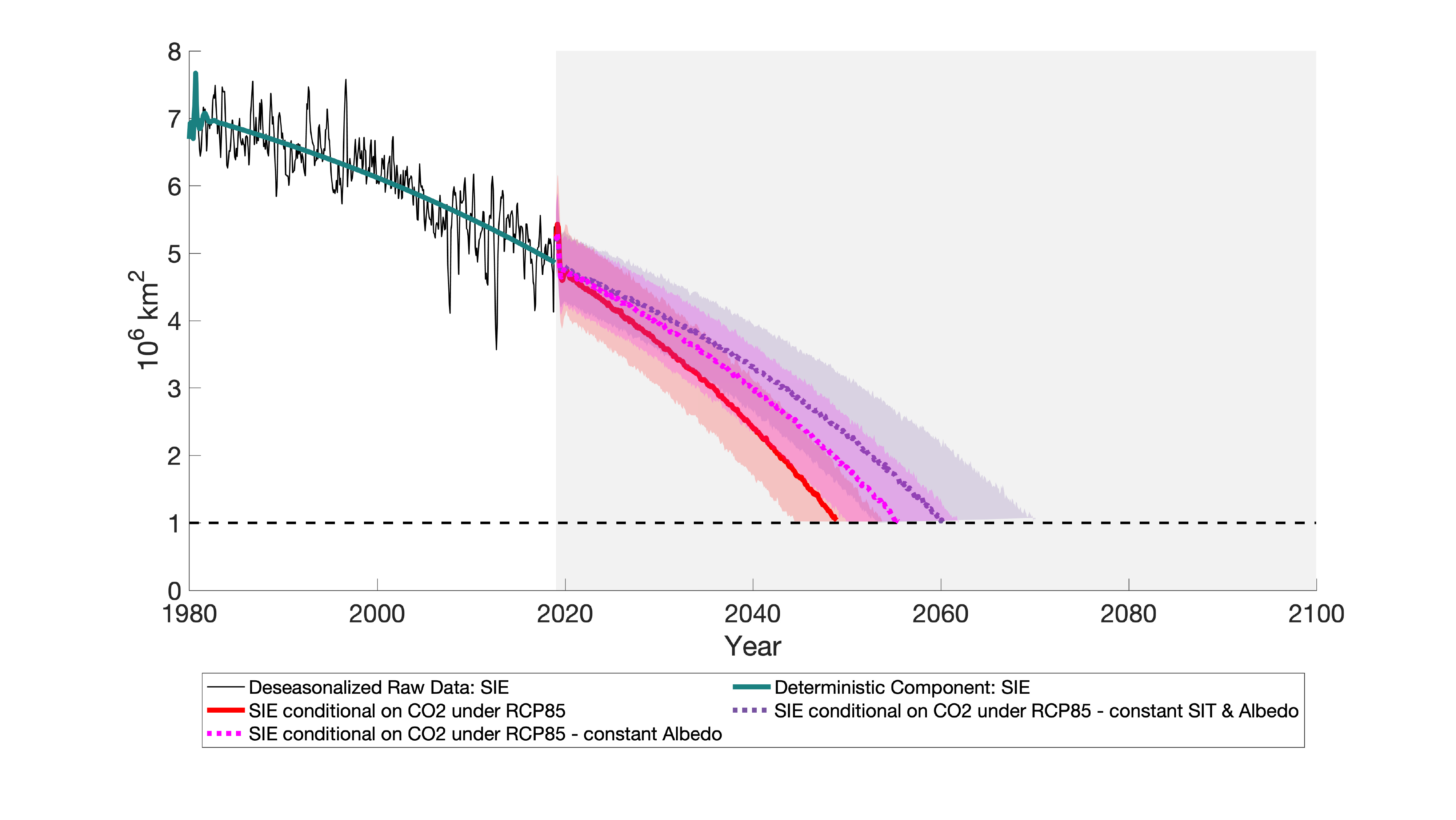}
\vspace*{-0.5cm}
\caption{RCP 8.5}\label{fig:TMA_dim8_Forecast_RCPs_a}
\end{subfigure}

\begin{subfigure}{\textwidth}
\centering
\includegraphics[scale = 0.35]{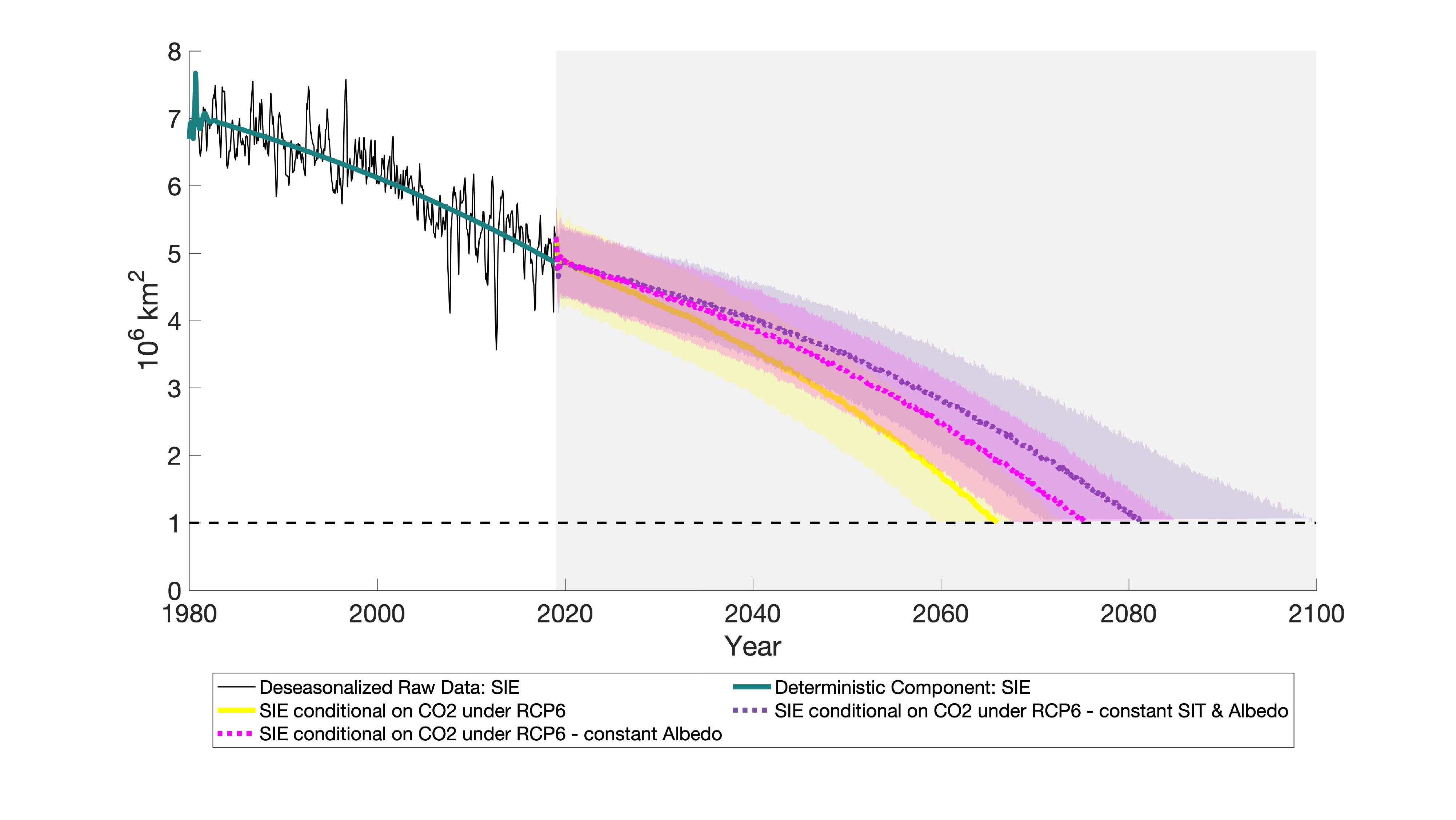}
\vspace*{-0.5cm}
\caption{RCP 6}\label{fig:TMA_dim8_Forecast_RCPs_b}
\end{subfigure}

\begin{subfigure}{\textwidth}
\centering
\includegraphics[scale = 0.35]{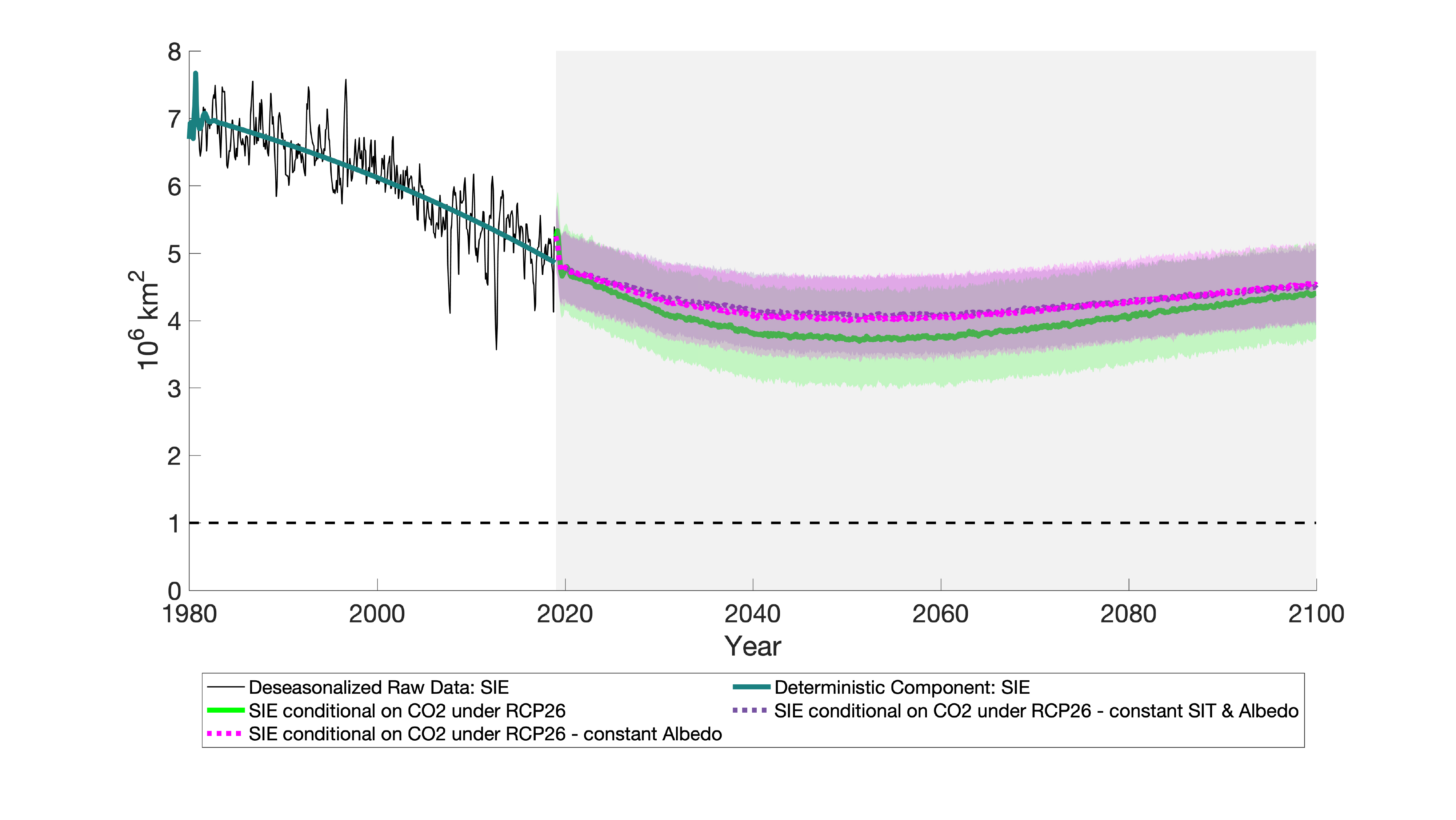}
\vspace*{-0.5cm}
\caption{RCP 2.6}\label{fig:TMA_dim8_Forecast_RCPs_c}
\end{subfigure}

\caption{Conditional Forecasts with and without Feedback}\label{fig:TMA_dim8_Forecast_RCPs}
\end{figure}

Figure \ref{fig:TMA_dim8_Forecast_RCPs} documents the corresponding results for \emph{RCP 8.5}, \emph{RCP 6} and \emph{RCP 2.6}. For each scenario, we show three different cases: (i) the projection of SIE under the respective RCP; (ii) the evolution of SIE under the respective RCP while keeping Albedo constant at its last in-sample deterministic value; (iii) the projection of SIE while keeping both Albedo and SIT constant at their last respective deterministic value. The latter are shown to be undeniable accelerants. First, fixing Albedo to its 2019 value and thus shutting down this particular long-run amplification effect postpones the date of reaching $1 \times 10^6$ $km^2$  by a bit less than a decade under both \emph{RCP 8.5} and \emph{RCP 6}. Arctic sea ice thickness plays a major role for the reaction and resilience of SIE to anthropogenic forcing. Figure \ref{fig:TMA_dim8_Forecast_RCPs} re-enforces this view by showing that preventing both SIT and Albedo from further decay could potentially postpone the zero-SIE event to the next century under \emph{RCP 6}. Under \emph{RCP 8.5}, shutting down both amplification channels starting from 2020 leads to SIE crossing the bar of $1 \times 10^6$ $km^2$  about a decade later.\footnote{The graphs are cut at the $1 \times 10^6$ $km^2$ bar as keeping SIT constant (which the thought experiment suggests) is untenable as SIE approaches 0: SIT cannot be constrained to be positive if SIE is 0.}  This feeds into the pictured non-linearity and acceleration of SIE loss and provides a potential justification for the finding in \cite{DieRu19} that a quadratic trend is a preferable approximation of long-run summer months' SIE evolution. 

\section{Conclusion and Directions for Future Research} \label{conclusion}

We proposed the VARCTIC as a middle ground alternative to purely theoretical or statistical modeling. It generates long-run forecasts that embody the interaction of many key variables without the inevitable opacity of climate models. First, we focus our attention on how the Arctic system responds to exogenous impulses and propagates them. Our results show that $CO_2$ anomalies have an unusually lasting effect on SIE which takes about a year to settle in. It is the only impulse that has the property of durably affecting SIE. Albedo and SIT are shown to play an important role in amplifying the response of SIE to $CO_2$ and AT shocks. In both cases, the conjunction of the two effects can double the cumulative impact of such shocks after two years.

Second, we focus on the systematic/deterministic part of the VARCTIC and conduct conditional forecasting experiments that again seek to quantify the effect of anthropogenic $CO_2$ and how feedback loops can amplify it. We condition on the future path of $CO_2$ and show that, within the context of our model, it is the prime source of uncertainty for the long-run forecast of SIE. \emph{RCP 8.5} implies 0 September SIE around 2054, \emph{RCP 6} says so around 2075 and finally, \emph{RCP 2.6} ($\sim$ Paris Accord) implies that such an event would never happen. We conclude the analysis by evaluating to which extent internal knock-on effects can amplify the long-run effect of $CO_2$ forcing. Our results provide statistical backing for the view that $CO_2$ shocks trigger feedbacks of other climate variables (as characterized here by Albedo and SIT), which substantially accelerate the speed at which SIE is headed toward 0.

There are many methodological extensions  within the VAR paradigm that could be of interest for future cryosphere research. For instance, Smooth-Transition VARs (with a popular application in \cite{AG2012}) could be used to accommodate for dynamics evolving over the seasonal cycle. Additionally, \cite{ScreenDeser2019} remark the importance of changing weather phenomena that transition through decadal cycles, such as the pacific oscillation. Time-varying parameters VARs could evaluate the quantitative relevance of such phenomena. Finally, some recent attention (\cite{chavas2019dynamic}) has been given to the potentially non-linear relationship between $CO_2$ and SIE. Methods that blend time series econometrics and Machine Learning of the like in \cite{GC_MRF} could reveal interesting insights on complex/time-varying relationships in the Arctic.

\clearpage

\bibliographystyle{ametsoc2014}
\bibliography{VARCTIC_references}

\clearpage
%
%
%
\appendix
\newcounter{saveeqn}
\setcounter{saveeqn}{\value{section}}
\renewcommand{\theequation}{\mbox{\Alph{saveeqn}.\arabic{equation}}} \setcounter{saveeqn}{1}
\setcounter{equation}{0}

\section{Appendix}

\subsection{Data Sources}

\begin{table}[H]
  \begin{threeparttable}
  \centering
  \scriptsize
  \caption{List of Variables}\label{tab:ListofAbb}
  \setlength{\tabcolsep}{0.1em}
    \begin{tabularx}{\textwidth}{sXs}
    \toprule \toprule
    \textbf{Abbreviation} & \textbf{Description} & \textbf{Data Source} \\
    \midrule
 \rowcolor{gray!15}
Age & Gridded monthly mean of \newline Sea Ice Age & 
\href{https://nsidc.org/data/nsidc-0611/versions/4}{EASE-Grid Sea Ice Age, Version 4} \\ 
   
AT & Gridded monthly mean of \newline Air Temperature & 
\href{https://www.esrl.noaa.gov/psd/data/gridded/data.ncep.reanalysis.surface.html}{NCEP/NCAR Reanalysis 1: Surface} \\ 
 
   \rowcolor{gray!15}   
 Albedo & Gridded monthly mean of \newline Surface Albedo & 
 \href{https://disc.gsfc.nasa.gov/datasets/M2TUNXRAD_5.12.4/summary?keywords=merra-2}{ MERRA-2} \\
      
$CO_2$ & Global monthly mean of $CO_2$ & 
\href{https://www.esrl.noaa.gov/gmd/ccgg/trends/gl_data.html}{NOAA  - Earth System Research Laboratories}\\

   \rowcolor{gray!15}  
LWGAB & Gridded monthly mean of \newline Surface Absorbed Longwave Radiation & 
\href{https://disc.gsfc.nasa.gov/datasets/M2TUNXRAD_5.12.4/summary?keywords=merra-2}{MERRA-2} \\

LWGEM & Gridded monthly mean of \newline Longwave Flux Emitted from Surface & 
\href{https://disc.gsfc.nasa.gov/datasets/M2TUNXRAD_5.12.4/summary?keywords=merra-2}{MERRA-2}  \\

   \rowcolor{gray!15}  
LWGNT & Gridded monthly mean of \newline Surface Net Downward Longwave Flux & 
\href{https://disc.gsfc.nasa.gov/datasets/M2TUNXRAD_5.12.4/summary?keywords=merra-2}{MERRA-2} \\

LWTUP & Gridded monthly mean of \newline Upwelling Longwave Flux at TOA & 
\href{https://disc.gsfc.nasa.gov/datasets/M2TUNXRAD_5.12.4/summary?keywords=merra-2}{MERRA-2}  \\

   \rowcolor{gray!15}  
PR & Gridded monthly mean of \newline Precipitation & 
\href{https://www.esrl.noaa.gov/psd/data/gridded/data.cmap.html}{CPC Merged Analysis of Precipitation (CMAP)} \\
  
SST & Median northern-hemispheric mean Sea-Surface \newline Temperature anomaly (relative to 1961-1990) & 
\href{https://www.metoffice.gov.uk/hadobs/hadsst3/data/download.html}{Met Office Hadley Centre} \\

   \rowcolor{gray!15}  
SIE & Gridded monthly mean of \newline Sea Ice Extent & 
\href{https://nsidc.org/data/g02135}{Sea Ice Index, Version 3} \\

SWGNT & Gridded monthly mean of \newline Surface Net Downward Shortwave Flux & 
\href{https://disc.gsfc.nasa.gov/datasets/M2TUNXRAD_5.12.4/summary?keywords=merra-2}{MERRA-2} \\

   \rowcolor{gray!15}  
SWTNT & Gridded monthly mean of \newline TOA Net Downward Shortwave Flux & 
\href{https://disc.gsfc.nasa.gov/datasets/M2TUNXRAD_5.12.4/summary?keywords=merra-2}{MERRA-2}  \\

TAUTOT & Gridded monthly mean of \newline In-Cloud Optical SIT of All Clouds &
\href{https://disc.gsfc.nasa.gov/datasets/M2TUNXRAD_5.12.4/summary?keywords=merra-2}{MERRA-2}  \\

   \rowcolor{gray!15}  
SIT & Gridded monthly mean of \newline Sea Ice Thickness & 
\href{http://psc.apl.uw.edu/research/projects/arctic-sea-ice-volume-anomaly/data/model_grid}{PIOMAS} \\

TCC & Gridded monthly mean of \newline Total Cloud Cover & 
\href{https://www.esrl.noaa.gov/psd/data/gridded/data.ncep.reanalysis.derived.otherflux.html}{NCEP/NCAR Reanalysis Monthly Means and Other Derived Variables} \\
 
    \rowcolor{gray!15}  
 TS & Gridded monthly mean of \newline Surface Skin Temperature & 
 \href{https://disc.gsfc.nasa.gov/datasets/M2TUNXRAD_5.12.4/summary?keywords=merra-2}{ MERRA-2}\\

\bottomrule \bottomrule
    \end{tabularx}%
\begin{tablenotes}[para,flushleft]
  Notes: The above series (before any transformation) are gathered in one file \href{https://philippegouletcoulombe.com/code}{here}.
  \end{tablenotes}
  \end{threeparttable}
  \end{table}

\clearpage

\subsection{Transmission Mechanism Analysis for a Shock to SIE}\label{sec:TMAtheory}
The purpose of the TMA analysis is to assess how the response of variable $i$ to a shock on variable $j$ changes, if a third variable $z$ were immune to the shock generated by variable $j$.
Here we follow \cite{Sims12} by differentiating between the \emph{direct} and \emph{indirect} effect. The former is variable $i$'s own response to the shock hitting variable $j$. However, the shock also affects variable $z$, which itself transmits the shock further to variable $i$. This channel is the \emph{indirect} effect of a shock to variable $j$ on the response of variable $i$. Hence, it is the latter that will explain the role of variable $z$ within the transmission channel of a shock to $j$ on $i$. To do so, \cite{Sims12} introduce \emph{artificial} shocks to variable $z$, which offset its own response to a shock to $j$. These \emph{artificial} shocks have two effects: (i) the IRF of variable $z$ will be zeroed over the whole IRF horizon; (ii)  the \emph{indirect} channel transmits the artificial shock onto variable $i$ and allows to identify the \emph{direct} effect of $j$ on $i$. 


This procedure requires the transformation of the structural VAR, given in equation \eqref{struct_VAR} into the reduced form VAR of equation \eqref{rf_VAR}, which reads as follows:

\begin{align} \label{rf_VAR_TMA}
\boldsymbol{y}_t = \boldsymbol{c} + \sum_{p=1}^{P}{A^{-1} \Psi_p}\boldsymbol{y}_{t-p} + A^{-1} \boldsymbol{\varepsilon}_t  \qquad ,
\end{align}

where $A^{-1}$ is the Cholesky decomposition of matrix $A$ in equation \eqref{struct_VAR}. This imposes the necessary restrictions in order to identify the contemporaneous relationships of the variables. In particular, it assumes higher ordered variables to have an immediate effect on variables that are ranked below, but not vice versa. As $CO_2$ is ordered first in all of our models, an exogenous shock to carbon dioxide in period $t$ will have an immediate effect on all of the other variables. The companion form of equation \eqref{rf_VAR_TMA} is
\begin{align} \label{rf_VAR_TMA_comp}
\boldsymbol{Y}_t = \boldsymbol{c} + \Phi \boldsymbol{Y}_{t-1} + A^{-1} \boldsymbol{\varepsilon}_t  \enskip,
\end{align}
where $\boldsymbol{Y}_t = \left[y_t \; y_{t-1} \; \cdots \; y_{t-p-1}\right]^\top$ and the corresponding companion matrix is

\begin{align}\label{eq:companion}
\Phi =  \begin{pmatrix}
    A^{-1} \Psi_1 & A^{-1} \Psi_2 & \cdots & \cdots & A^{-1} \Psi_p \\ 
    I                     &  0					&		0	  &	\cdots & 0  \\ 
    0                    &  I					    &		0	  &	\cdots & 0  \\
    \vdots 			   &  \vdots			&  \ddots & \vdots & \vdots \\
     0                    &  \cdots			& \cdots &	I   		& 0 
  \end{pmatrix} \enskip .
\end{align}

An equivalent way (to what laid out in section \ref{sec:irf}) of constructing IRFs, i.e. the response of variable $i$ to a structural shock on variable $j$ over all horizons $h = 0, ..., H$, is to proceed iteratively. Hence, for a given period $h$, the response of $i$ to a shock hitting $j$ is given by
\begin{align}\label{IRF_VARP}
IRF(j\rightarrow i, h) = e_i 	\Phi^{h} A_{\bullet,j}^{-1} 
\end{align}
where $e_i$ is a selection vector of dimension $1 \times M$ with 1 at entry $i$ and 0 otherwise. $A_{\bullet,j}^{-1}$ elicits the $j^{th}$ column of $A^{-1}$. Following \cite{Sims12}, switching off the \emph{indirect} effect of a shock to variable $j$ on $i$ via variable $z$ amounts to $IRF(j\rightarrow z, h) = 0 \; \forall \; h = 0,...,H$. That requires the \emph{artificial} shocks, $\varepsilon_{z,h}$, to be calibrated such that the response of variable $z$ to a shock to variable $j$ is zero over the whole IRF period. Hence, at $h = 1$ the \emph{artificial} shock $\varepsilon_{z,1}$ is

\begin{align} \label{TMA_art_1_1}
\varepsilon_{z,1}= - \frac{A^{-1}_{j,1}}{A^{-1}_{z,1}}		\enskip.
\end{align}

\noindent As these shocks are transmitted through time, the \emph{artificial} response $\varepsilon_{z,h}$ has to account for all the past shocks, $\varepsilon_{z,h-1}$, for any periods beyond $h = 1$:

\begin{align} \label{TMA_art_h_1}
\varepsilon_{z,h}= - \frac{IRF(j\rightarrow z, h) + \sum^{h-1}_{h'=0} e_z \Phi^{h-h'} A^{-1}_{\bullet,j} \varepsilon_{z,h'}} {e_z A^{-1}_{z}}		\enskip .
\end{align}

\noindent The altered IRFs (that omits the transmission channel $z$) for all the variables in the model to a shock to $j$ is
\begin{align} \label{TMA_IRF_manip}
IRF_{-z}(j\rightarrow \boldsymbol{i}, h) = IRF(j\rightarrow \boldsymbol{i}, h) + \sum^{h}_{h'=0} e_z \Phi^{h-h'} A^{-1}_{\bullet,j} \varepsilon_{z,h'}	\enskip.
\end{align}

So far, we have reviewed how IRF decomposition works when one is interested in shutting down a single channel at a time. In contrast to \cite{Sims12}, our VAR comprises more than three variables. Therefore, in some cases, it is desirable to shut-down not only one, but a group $Z \in M \setminus \left\lbrace i,j \right\rbrace $ of \emph{indirect} channels. To do so, equations \eqref{TMA_art_1_1} and \eqref{TMA_art_h_1} need to be generalized. At impact, the \emph{artificial} response of variable $z$ to a shock to $j$ does not only have to offset the \emph{direct} effect of $j$, but also the \emph{indirect} effect of a shock $j$ via the \emph{indirect} effect of all the other \emph{artificial} responses ($\boldsymbol{\varepsilon^{+}_{z^{+},1}}$) of those variables in $Z$ which are ordered above $z$.\footnote{$z^{+}$ denotes all those variables in $Z$ which are ordered above $z$.} This amounts to the following extension of equation \eqref{TMA_art_1_1}:

\begin{align} \label{TMA_art_1_Z}
\varepsilon_{z,1}= - \left(\frac{A^{-1}_{j,1}}{A^{-1}_{z,1}}	+ \frac{\sum_{m \in z^{+}} \varepsilon_{m,1}}{A^{-1}_{z,1}} \right)	\enskip .
\end{align}

\noindent Also equation \eqref{TMA_art_h_1} has to be adjusted accordingly. However, at horizons $h > 1$ the \emph{artificial} response $\varepsilon_{z,h}$ will not only have to offset the contemporaneous effects of $z^{+}$, but also compensate for the \emph{artificial} responses of all other variables in $Z$ over the period $h' = 0 \; \cdots \; h-1 $:

\begin{align} \label{TMA_art_h_Z}
\varepsilon_{z,h}= - \frac{IRF(j\rightarrow z, h) + \sum^{h-1}_{h'=1} e_z \Phi^{h-h'} A^{-1}_{j} \varepsilon_{z,h'} + \sum^{h-1}_{h'=1} \sum_{n \in Z} \varepsilon_{n,h'} + \sum_{m \in z^{+}} \varepsilon_{m,1}} {e_z A^{-1}_{z}}		\enskip .
\end{align}

\noindent Equation \eqref{TMA_IRF_manip} for the modified IRF ($IRF_{-z}(j\rightarrow \boldsymbol{i}, h)$) remains intact.

\clearpage

\subsection{Bayesian Estimation Details} \label{sec:priors}

Bayesian inference implies the use of priors, which degree of informativeness is usually determined by the user. To be as agnostic as possible, we use the technique of \cite{giannone2015prior} to choose the tightness of priors as to optimally balance bias and variance in a data-driven way.\footnote{Setting priors' tightness in such a way can be understood as analogous (at a philosophical level) to setting tuning parameters using cross-validation in Machine Learning.} The prior structure, however, must be chosen. We estimate our benchmark Bayesian VARCTIC with a standard Minnesota prior. In this simple framework,  $\bm{\Sigma}$, the variance-covariance matrix of the VAR residuals, is treated as known.\footnote{This choice is motivated by the fact that it facilitates the optimization of hyperparameters. As it turns out, optimizing tuning parameters has more impact on resulting IRFs and their respective credible regions than treating $\bm{\Sigma}$ as unknown, when using for instance an Independent Normal Wishart (with Gibbs sampling).}  Thus, the remaining parameters of the model reduce to the vectorized matrix $\beta = vec\left(\left[ \bm{\Phi_1} \cdots \bm{\Phi_p} \; \textbf{c} \right]^\top \right)$ 
of dimension $ (M^2p + M) \times 1$. The posterior distribution of $\beta$, $\pi\left(\beta|y\right)$, is obtained by the product of the likelihood function of the data $f\left(y|\beta\right)$, and the prior distribution of $\beta$, $\pi\left(\beta\right)$. Hence, by sampling from the posterior distribution  
 $\pi\left(\beta|y\right) \propto f\left(y|\beta\right) \pi\left(\beta\right)$
we can quantify both the uncertainty around $\beta$, but also more interesting transformations of it, such as IRFs and forecasts.\footnote{In the latter case, the credible region will naturally comprehend the uncertainty from the act of recursive forecasting itself, but also the fact that it relies on unknown parameters that must be estimated.}  The prior distribution for $\beta$ is the multivariate normal distribution $\pi\left(\beta\right) \sim N\left(\beta_0,\Sigma_0\right)$. The Minnesota prior is a specific structure for values of both $\beta_0$ and $\Sigma_0$.\footnote{As a reference, a Ridge regression would imply $\beta_0=\boldsymbol{0}$ and $\Sigma_0$ being a diagonal matrix with identical diagonal elements.} In words, it allows concisely to parameterize heterogeneity in both the prior mean and variance. It consists of three major elements: the first one is about $\beta_0$ and the last two concern $\Sigma_0$.
\begin{enumerate}
\item For any equation $y_{m,t}$ with $m = 1,...M$ -- where $M$ is the total number of observed variables in the VAR -- all parameters are shrunk to 0 except for its first \textit{own} lag $y_{m,t-1}$. The latter is usually shrunk to a value $b_{AR}$ between 0.5 and 1. This can be interpreted as shrinking each VAR equation to the much simpler and parsimonious AR(1) process. Given the persistent nature of time-series data, this structure for $\beta_0$ is much more appropriate than that of Ridge regression (or LASSO), which shrinks all coefficients homogeneously towards 0.
\item  It is often observed in multivariate time series models that $y_{m,t-1} \rightarrow y_{m,t}$ will be way stronger than almost any of the $y_{\tilde{m},t-1} \rightarrow y_{m,t}$ (with $\tilde{m} \neq m$) relationships. $\lambda_2$ therefore calibrates the relative intensity of shrinking dynamic cross-correlations versus that of autocorrelations.
\item Distant partial lag relationships (say $y_{m,t-12} \rightarrow y_{m,t}$) are expected to be of smaller magnitude than close ones like $y_{m,t-1} \rightarrow y_{m,t}$, and $y_{m,t-2} \rightarrow y_{m,t}$. $\lambda_3$ determines the additional intensity of distant lags shrinkage.
\end{enumerate}

The overall tightness of the prior apparatus is determined by $\lambda_1$.\footnote{For further details, explicit mathematical formulation of the prior and additional discussion on priors for VARs, the reader is referred to \citep{BearMatlab}.} To draw a parallel to penalized regression (like Ridge and LASSO), a small $\lambda_1$ in a Bayesian VAR increases regularization in a way analogous to increasing the $\lambda_{RIDGE}$ -- that is, by pushing the BVAR estimate $\hat{\Phi}$ away from $\hat{\Phi}_{OLS}$. Following \cite{giannone2015prior}, we optimize/estimate hyperparameters within some grid and  the total number of posterior draws is 2000. The optimal values for VARCTIC 8 are
$\{ b_{AR}, \lambda_1, \lambda_2, \lambda_3 \}=\{0.9, 0.3, 0.5, 1.5\}$. We fix the number of lags in VARCTIC 8 to $P=12$ and to $P=3$ in VARCTIC 18 respectively.

Finally, given the very smooth look of deseasonalized $CO_2$ in Figure \ref{fig:20Vars_ds}, one could worry that it merely acts as a proxy for an omitted linear trend. We view the use of trends as undesirable in our multivariate setup as it would undermine the capacity of the VARCTIC to be a "complete" model. Including a trend would make it rely on an unknown/unexplained latent force -- which is at odds with the main goal of our modeling strategy. For the sake of completeness, we nevertheless estimate such models to find out that the inclusion of an exogenous time trend is in fact not preferred by the data according to the Deviance Information Criterion (DIC, a generalization of the well-known AIC). VARCTIC 8 has a DIC of -6894.35 and adding an exogenous trend makes it -6817.32. The smallest value being preferred, this justifies on a data-driven basis the exclusion of the trend. While seemingly technical of point, this means the VARCTIC 8 system, based solely on dynamic relationships of observable data, can generate/simulate the observed SIE downward path.

\clearpage

\subsection{Different Ordering} \label{sec:altord}
In this section, we check the sensitivity of the responses of SIE to a shock of any of the other variables when varying the ordering of variables compared to the benchmark VARCTIC 8 in section \ref{sec:results}. The priors and lags remain unaltered to the specification outlined in section \ref{sec:priors}. The ordering now reads: $CO_2$, AT, SST, TCC, PR, SIE, SIT, Albedo.


\begin{figure}[H]
\caption{IRFs: Response of Sea Ice Extent} \label{fig:SIE_8dim_IRF_DO_V1}
\centering

\includegraphics[width = \textwidth, trim=1cm 1cm 1cm 3cm, clip]{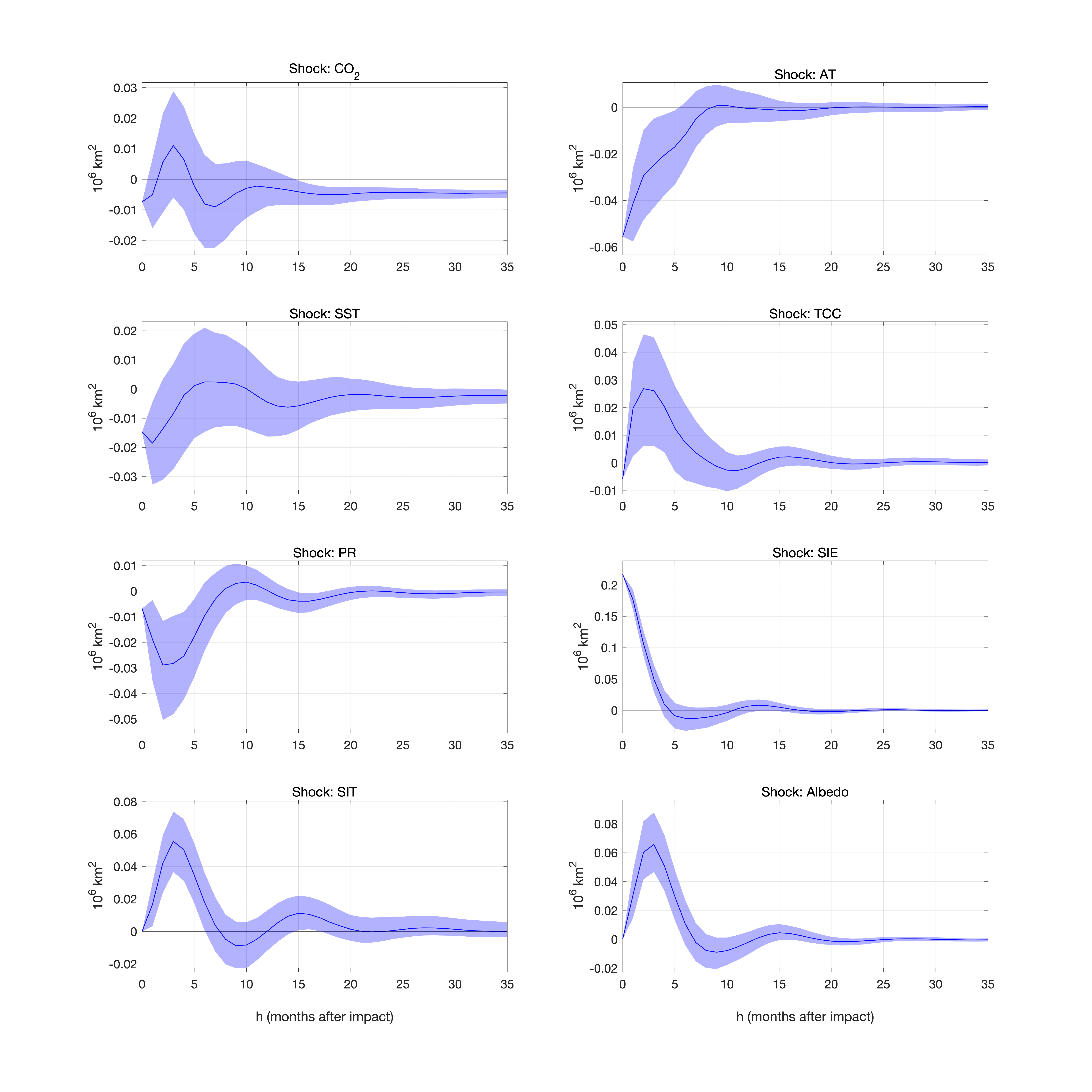}
\end{figure}

A comparison of the responses of the benchmark VARCTIC 8 in Figure \ref{fig:SIE_8dim_IRF} and the IRFs after reordering the model (Figure \ref{fig:SIE_8dim_IRF_DO_V1}) documents the robustness of results to different identification schemes. A second -- more radical -- variation in the model set-up locates Albedo at position two. Hence, a shock to Albedo will contemporaneously affect all the other variables except $CO_2$: $CO_2$, Albedo, TCC,	PR, AT, SST, SIE, SIT.


\begin{figure}[H]
\caption{IRFs: Response of Sea Ice Extent} \label{fig:SIE_8dim_IRF_DO_V2}

\centering
\includegraphics[width = \textwidth, trim=1cm 1cm 1cm 3cm, clip]{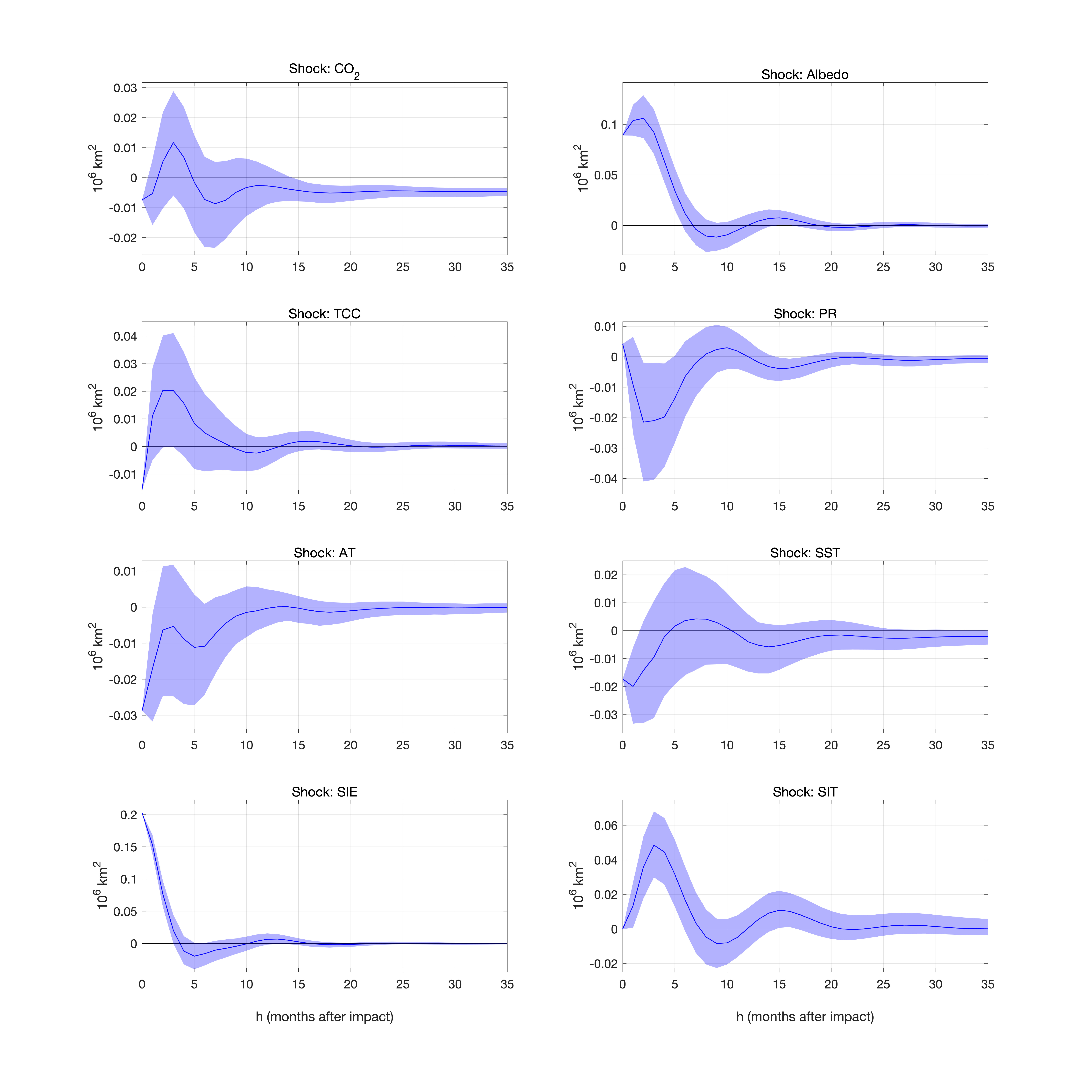}
\end{figure}

For most of the effects, the shapes remain robust in comparison with Figure \ref{fig:SIE_8dim_IRF}. Only the response to air temperatures deviates visibly with the statistically significant impact in the short-run now vanishing.

\clearpage

\subsection{Results of VARCTIC 18} \label{sec:v18}

VARCTIC 18, including all the variables in Table \ref{tab:ListofAbb}, tests the robustness of VARCTIC 8 projection of SIE. The ordering of variables in VARCTIC 18 reads as follows: SWGNT, SWTNT, $CO_2$, LWGNT, TCC, TAUTOT, PR, TS, AT, SST, LWGAB, LWTUP, LWGEM, SIE, Age, SIT, EMIS and Albedo. Due to the increased number of variables, the lags were reduced to 3 and the estimation period starts in 1984 due to some series unavailability. With more parameters to estimate, the prior specification slightly tightens to $\{ b_{AR}, \lambda_1, \lambda_2, \lambda_3 \}=\{0.8, 0.5, 0.5, 3\}$.

\begin{figure}[t]
\centering
\includegraphics[scale=0.35]{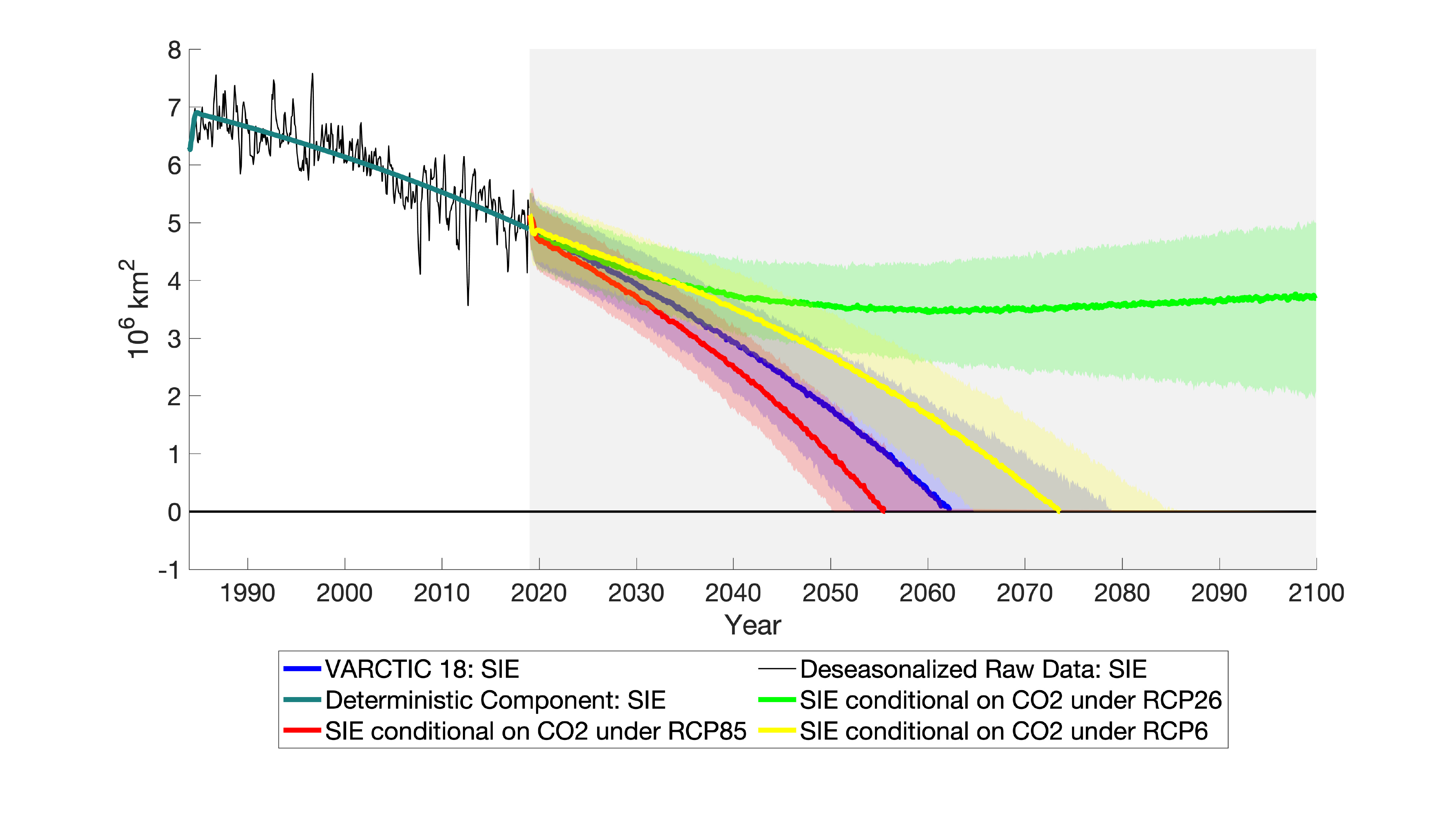}\\
\caption{Evolution of SIE different scenarios of $CO_2$ in  VARCTIC 18}\label{fig:18BVAR_$CO_2$_UNCOND_RCP85_RCP26_DD}
\end{figure}

The forecasts of SIE under the specification of VARCTIC 18 are all reported in Figure \ref{fig:18BVAR_$CO_2$_UNCOND_RCP85_RCP26_DD}. This includes both the unconditional forecast and those conditional on \textit{RCP}'s.
The median unconditional VARCTIC 18 forecast a blue Arctic in September 2062, which is in the very neighborhood of VARCTIC 8. This result suggests VARCTIC 8 to comprise the key variables for a proper and robust long-run projection of Arctic sea ice.  The projected ice-free dates under the \emph{RCP 6} and \emph{RCP 8.5} scenarios are also consistent with the results reported by VARCTIC 8 in Figure \ref{fig:8BVAR_Projections_DiffRCPs_CO2}. The trajectory of SIE under \emph{RCP 2.6}, however, slightly changes and seems to stabilize rather than recover by the end of the century.

\begin{figure}[t]
\centering
\includegraphics[width=\textwidth, trim=0cm 0cm 0cm 0cm, clip]{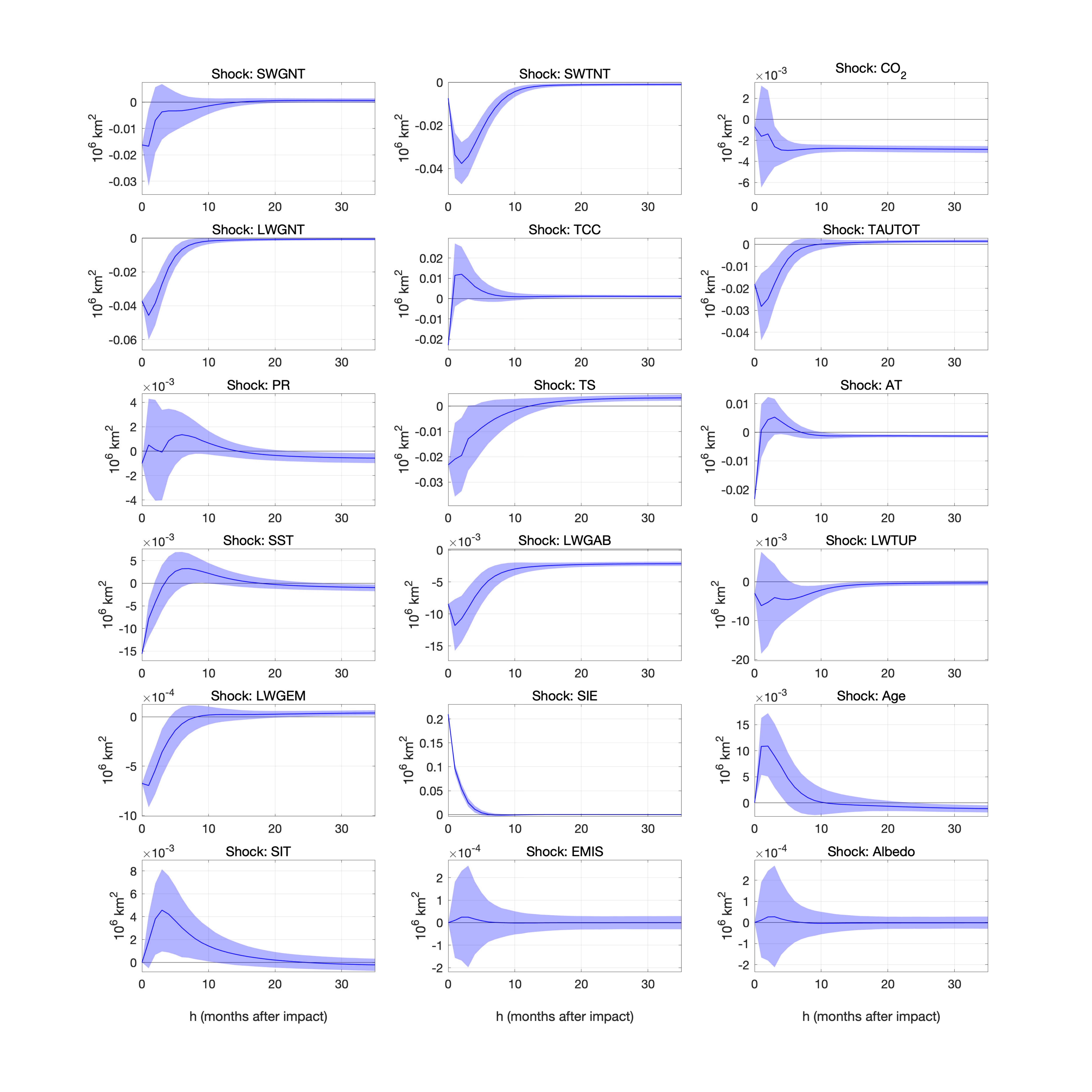}\\
\caption{IRFs: Response of Sea Ice Extent in VARCTIC 18. Shade is the 90\% credible region.}\label{fig:SIE_18dim_IRF}
\end{figure}

The IRFs of SIE are shown in Figure \ref{fig:SIE_18dim_IRF}. Those of key variables (as included in VARCTIC 18) remain roughly unchanged in VARCTIC 18. Most interestingly, in VARCTIC 18, not only $CO_2$ shock has the effect of triggering a durably decreasing SIE, but also LWGAB, which measures the longwave radiation absorbed by the surface and AT. Many other shocks have significant impacts in the short run but only $CO_2$, LWGAB, and AT shocks have the unique property of durably pushing the system out of the former equilibrium. 

\clearpage

\subsection{Stochastic De-seasonalization}\label{sec:ssm_theory}

%

As a robustness check, we verify that our main results hold if we employ a radically different technique to take out seasonality. In this subsection, we adopt the approach of structural time series (\cite{Harvey90} and \cite{harvey2014}) where $y^{raw}$ is split into three somewhat intuitive parts:
\begin{align*}
&		y_t^{raw}  = \mu_t +  \gamma_t +  \eta_t
		\end{align*}
a trend component $\mu_t$; a seasonality component $ \gamma_t $ and a  (possibly autocorrelated) noise component $\eta_t$. Each of them is stochastic and has its own law of motion. The structure and law of motions we use follow the well-established Harvey Basic Structural Model \citep{harvey1983}. The model reads as follows:

		\begin{align*}
& \mu_t  = \mu_{t-1}+ \beta_t + u_t \\
& \beta_t  = \beta_{t-1} + v_t 
		\end{align*}
\begin{align*}
& \gamma_t = -\sum_{m=1}^{11} \gamma_{t-m} + w_t
		\end{align*}
		\begin{align*}
		\left( \eta_t, u_t, v_t, w_t \right) 
~ \sim ~ iid \, N \left (0 , ~ \Sigma \right )
		\end{align*}
		\begin{align*}
\Sigma = 
\begin{pmatrix}
\sigma^2_{\eta \eta} &0 & 0 &0\\
0& \sigma^2_{uu} &0 &0\\
0&0 &\sigma^2_{vv} &0\\
0&0 &0 &\sigma^2_{ww} \\
\end{pmatrix}
		\end{align*}
		
\begin{figure}[!htb]
\captionsetup{justification=centering}
\caption{Basic Structural Model: 8 Variables \\ \emph{Extracted Trends adjusted for average September-Seasonality}} \label{fig:8Vars_ssm}
\centering
\includegraphics[width=\textwidth]{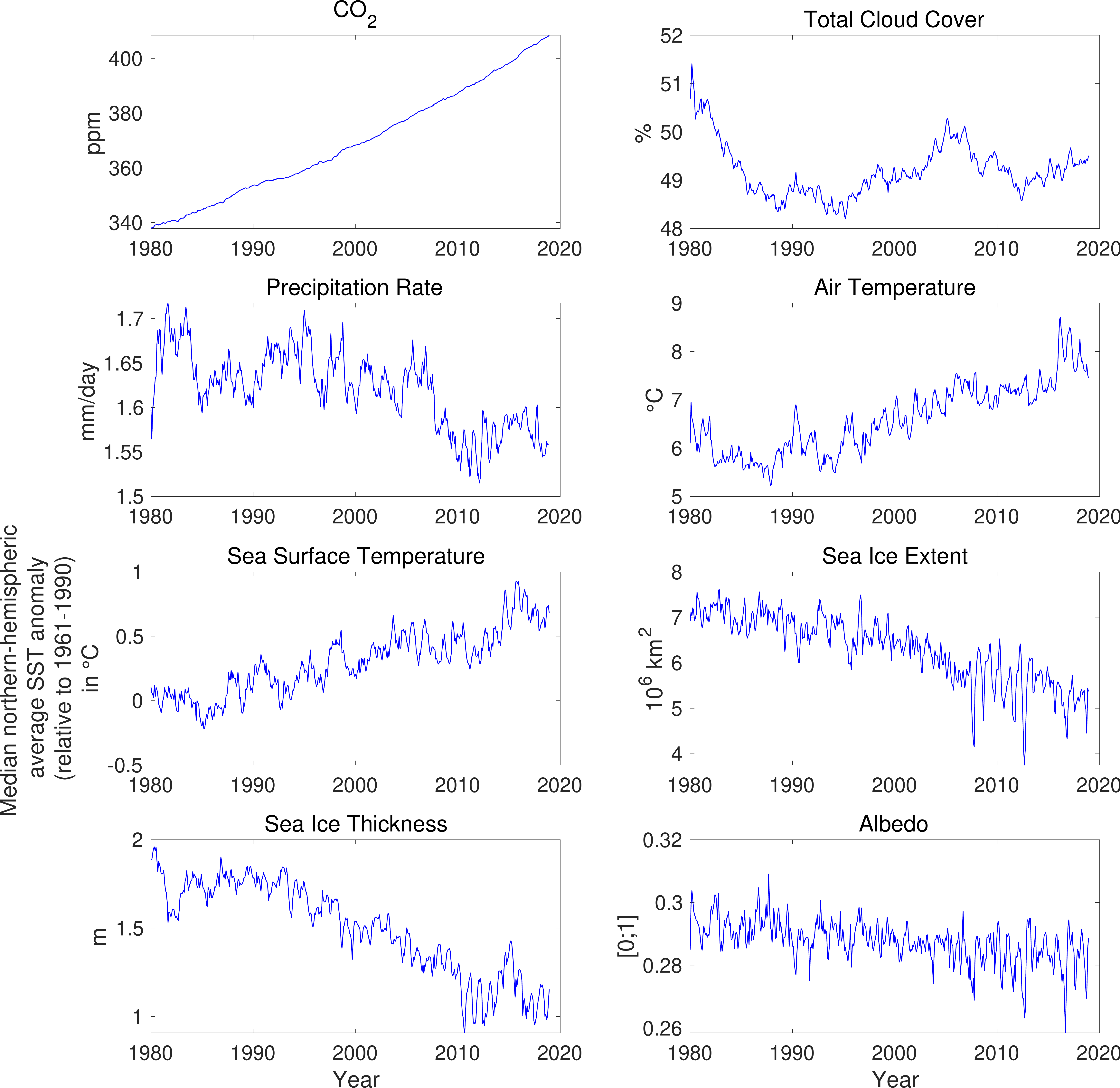}
\end{figure}

The law of motion is that of \cite{harvey1983} and fits in a traditional state space model. The trend $\mu_t$ is a random walk with a stochastic drift. The drift $\beta_t$ is itself evolving according to a random walk. For instance, this means that $\mu_{SIE,t}$, the trend of SIE, is trending down stochastically at a rate $\beta_{SIE,t}$. That (negative) growth rate is itself allowed to evolve. A quick look at a flexibly modeled trend of SIE suggests that allowing for a time-varying growth rate is necessary given the acceleration and deceleration of SIE melting in the 2000's. Figure \ref{fig:8Vars_ssm} shows the complete set of stochastic trends resulting from the BSM.

The extraction of trends as a first step and their subsequent modeling as a second step is analogous to standard practice in macroeconomics, but not similar. In macroeconomics, it is customary in a strand of empirical work to filter the data as a pre-processing step. The VAR is then estimated on the extracted cycles, which is simply the difference of the raw data and the estimated trend. Here, we are indeed doing the filtering step first but using trend components -- rather than seasonality and short-run noise -- for the second step. However, our trend components $\mu_t$ are rather stochastic with respect to what is usually seen in economics.


\subsubsection{The Benchmark Specification and Results}

Following \cite{giannone2015prior}, we obtain the optimal hyperparameters:
\begin{itemize}
\item Autoregressive Coefficient: = 1;
\item Overall tightness is $\lambda_1$ = 0.3;
\item Cross-variable weighting is $\lambda_2$ = 0.5;
\item Lag decay is $\lambda_3$ = 1.51;
\item Exogenous variable tightness: $\lambda_4$ = 100;
\end{itemize} 

The date of the zero-lower bound of the \emph{stochastic de-seasonalized version} remains in the neighborhood of the benchmark model. In this specification, the Arctic would be ice-free by the year 2061.

\begin{figure}[!h]
\caption{Trend Sea Ice Extent \\ \emph{Stochastic De-seasonalization}}\label{fig:SIE_8dim_ssm}
\centering
\includegraphics[scale=0.35]{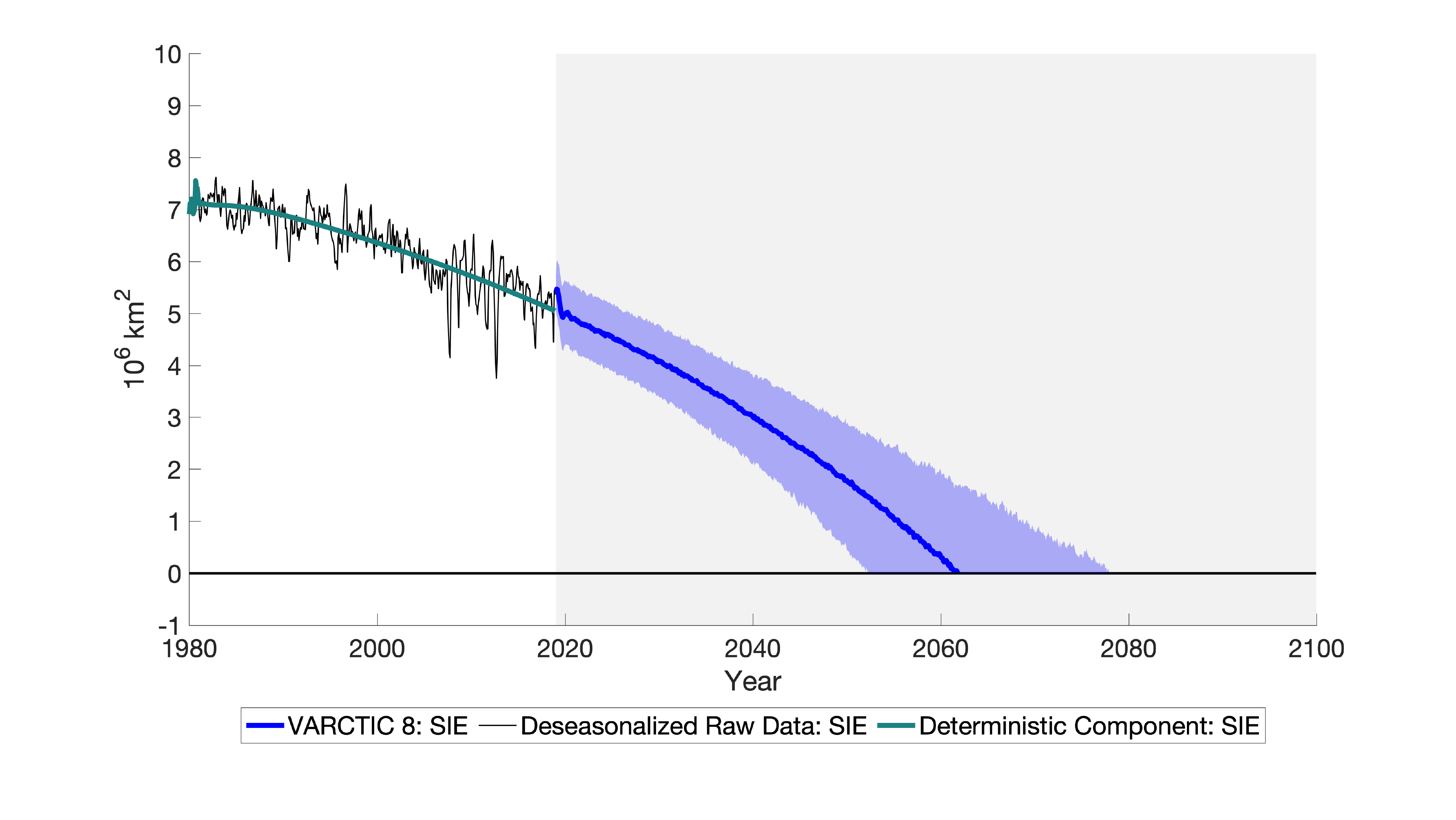}
\end{figure}

\begin{figure}[!h]
\caption{IRFs: Response of Sea Ice Extent \\ \emph{Stochastic De-seasonalization}} \label{fig:SIE_8dim_IRF_ssm}
\centering
\includegraphics[width = \textwidth, trim=1cm 1cm 1cm 3cm, clip]{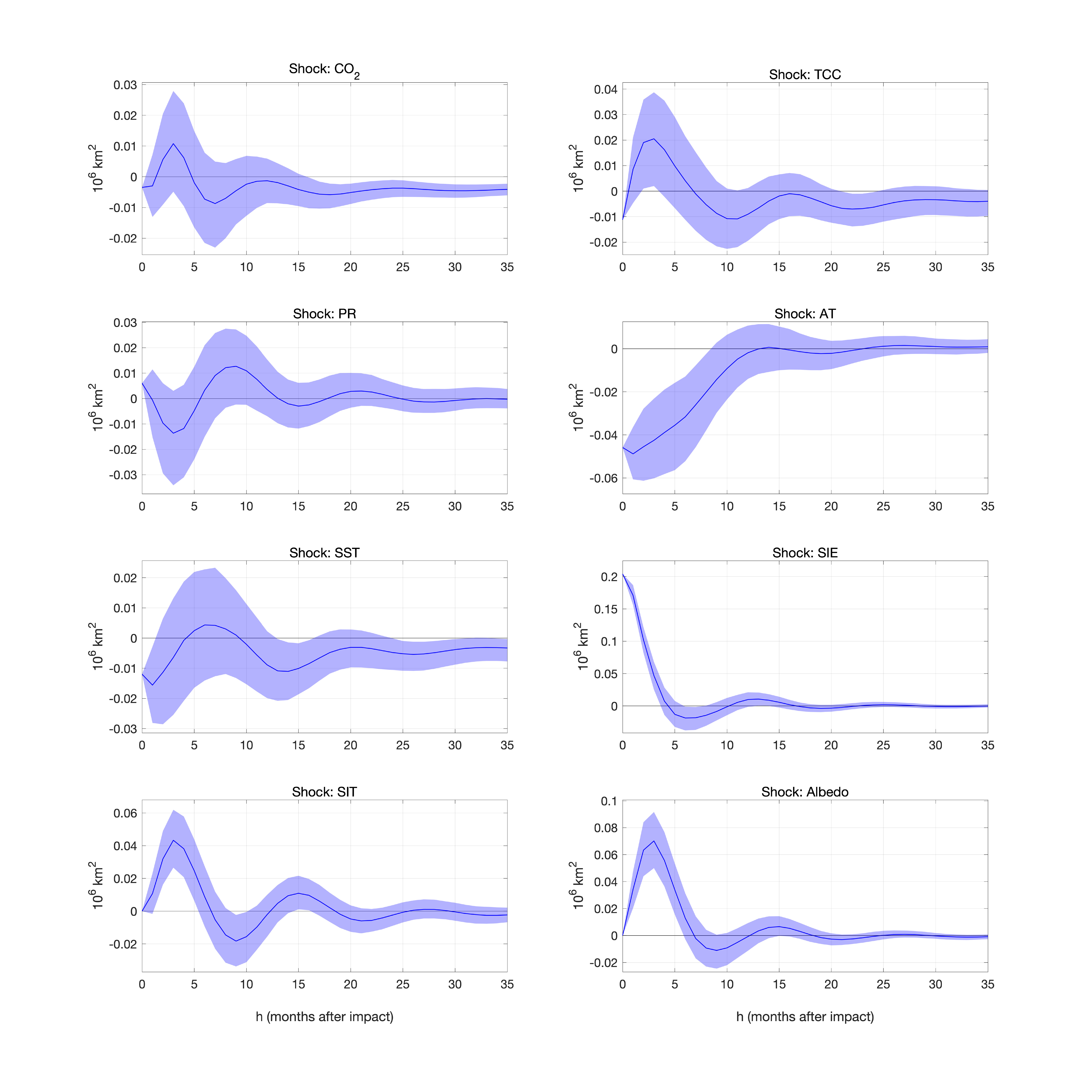}
\end{figure}

\begin{figure}[!h]
 \captionsetup{justification=centering}
\caption{Evolution of SIE under different Scenarios of $CO_2$ \\ \emph{Different Scenarios} \\ \emph{Stochastic De-seasonalization - Extracted trend adjusted for mean September-seasonality}} \label{fig:8BVAR_$CO_2$_UNCOND_RCP85_RCP26_ssm}
\centering
\includegraphics[scale=0.35]{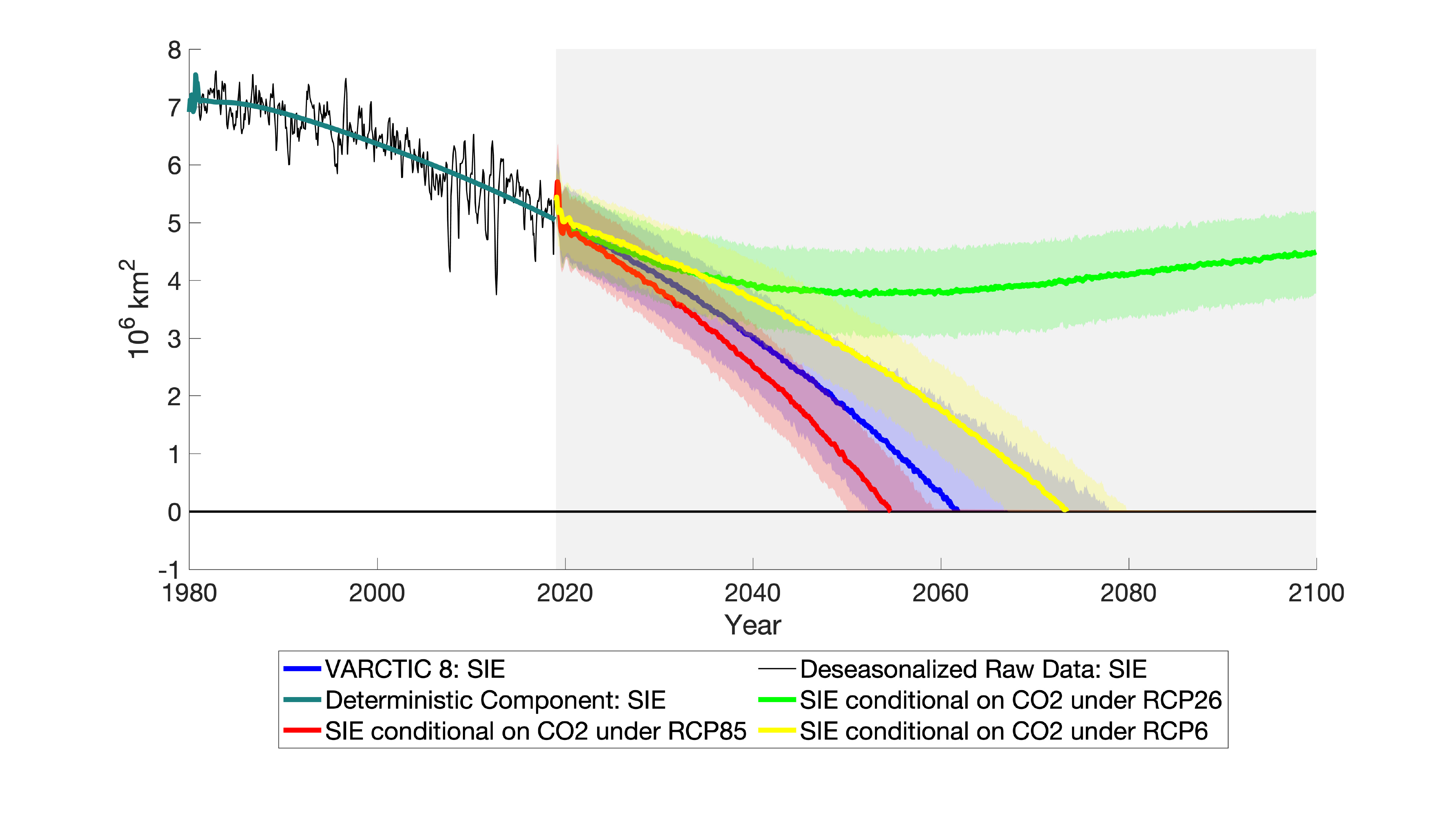}
\end{figure}

\begin{figure}[!h]
 \captionsetup{justification=centering}
\caption{Evolution of SIE under different Scenarios of $CO_2$ \\ \emph{Different Scenarios} \\ \emph{Stochastic De-seasonalization: \\ Extracted trend adjusted for yearly September- \& March-seasonality}} \label{fig:8BVAR_$CO_2$_UNCOND_RCP85_RCP26_ssm_SepMarch_y}
\centering
\includegraphics[scale=0.35]{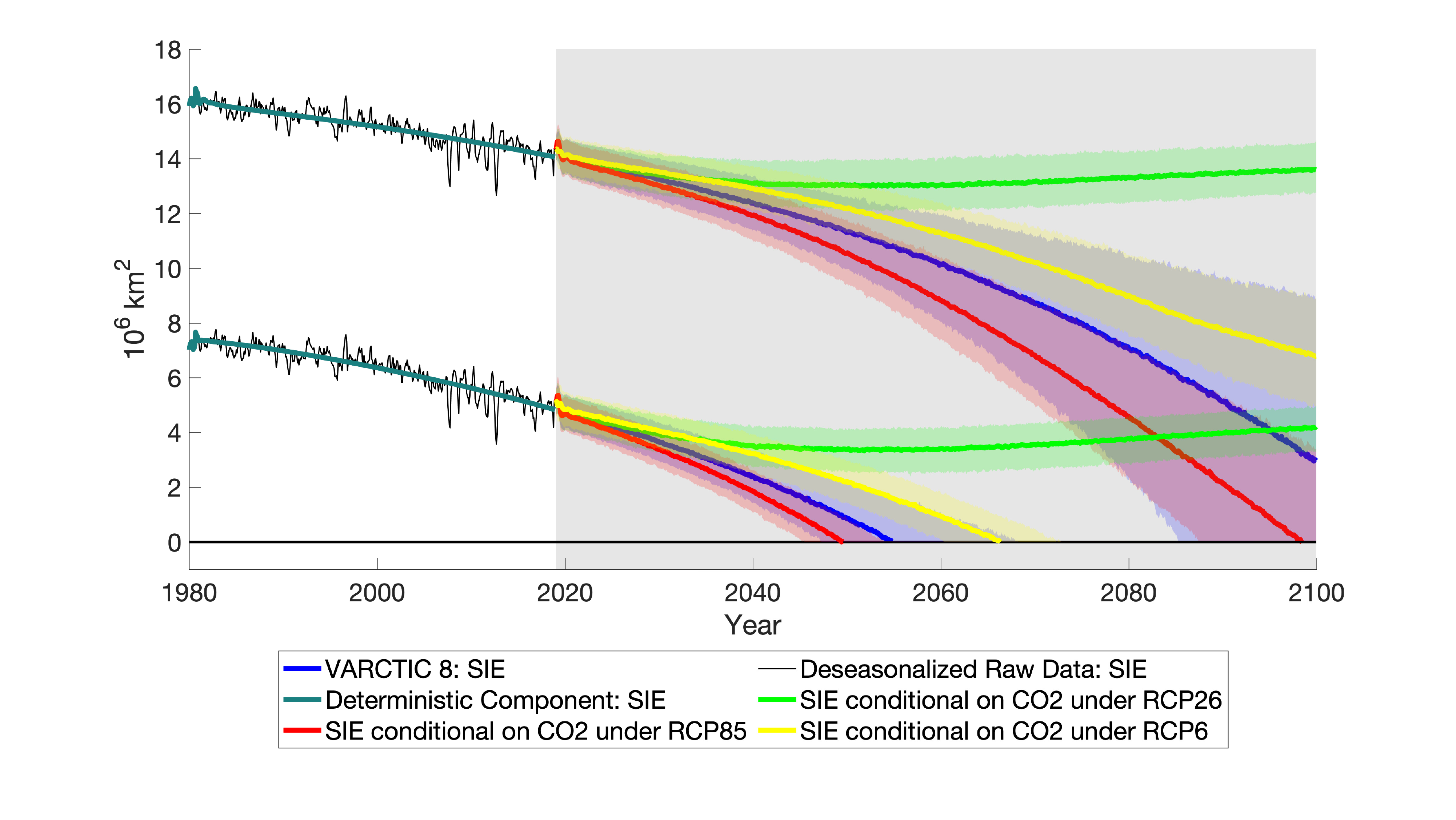}
\end{figure}

As the BSM specification allows for evolving seasonality, we can also use it to obtain more flexible month-specific VARCTICs. The benchmark specification implies that we can transform our series into a string of "synthetic" Septembers or Marchs by simply adding or subtracting a constant. In the evolving seasonality model, one can rewrite a slowly widening seasonal pattern as the expression of heterogeneous trends across seasons. Thus, rather than adding back the mean (over time) of $\gamma_{t,September}$ to $\mu_t$ to fit the model on static synthetic Septembers, we can add back $$\tilde{\gamma}_{t,September} = \sum_{t'=1}^{T} I(t' = t){\gamma}_{t',September}$$  to model evolving synthetic Septembers (or any month of interest). Unlike our benchmark specification, this approach allows for summer vs non-summer months to have different trends. Figure \ref{fig:8BVAR_$CO_2$_UNCOND_RCP85_RCP26_ssm_SepMarch_y} reports results of our conditional forecasting analysis conducted for two radically different months. While March's SIE is linearly trending in-sample, the projections suggest a potential acceleration of melting in the second half of the century -- with widening uncertainty.

\end{document}